\documentclass[journal=jacsat,keywords=true,manuscript=article]{achemso}
\setkeys{acs}{etalmode=truncate,maxauthors=0,usetitle=true}
\usepackage{graphicx}
\usepackage{amsmath}
\usepackage{caption}
\usepackage{subcaption}
\usepackage{color}
\usepackage{appendix}
\usepackage{rotating}
\usepackage{threeparttable}
\usepackage{multirow}
\usepackage{longtable}
\usepackage{mathrsfs}
\usepackage{xr}
\usepackage{bigints}
\usepackage[final]{pdfpages}
\newcommand{\lpfrev}[1]{\textcolor{black}{#1}}
\usepackage{atbegshi}% http://ctan.org/pkg/atbegshi
\usepackage[version=4]{mhchem}
\usepackage{pdfpages}

\author{Pengfei Li}
\affiliation{State Key Laboratory of Precision Spectroscopy, School of Physics and Materials Science, East China Normal University, Shanghai 200062, China}
\author{Xiangyu Jia}
\affiliation{State Key Laboratory of Precision Spectroscopy, School of Physics and Materials Science, East China Normal University, Shanghai 200062, China}
\altaffiliation{Current address: NYU-ECNU Center for Computational Chemistry at NYU Shanghai, Shanghai 200062, China}
\author{Xiaoliang Pan}
\affiliation{Department of Chemistry and Biochemistry, University of Oklahoma, Norman OK 73019, United States of America}
\author{Yihan Shao}
\affiliation{Department of Chemistry and Biochemistry, University of Oklahoma, Norman OK 73019, United States of America}
\author{Ye Mei}
\email{ymei@phy.ecnu.edu.cn}
\affiliation{State Key Laboratory of Precision Spectroscopy, School of Physics and Materials Science, East China Normal University, Shanghai 200062, China}
\alsoaffiliation{NYU-ECNU Center for Computational Chemistry at NYU Shanghai, Shanghai 200062, China}
\alsoaffiliation{Department of Chemistry and Biochemistry, University of Oklahoma, Norman OK 73019, United States of America}

\title[]{Accelerated Computation of Free Energy Profile at \textit{ab initio} Quantum Mechanical/Molecular Mechanics Accuracy via a Semi-Empirical Reference-Potential. I. Weighted Thermodynamics Perturbation}
	
\begin{document}
	
\begin{abstract}
Free energy profile (FE Profile) is an essential quantity for the estimation of reaction rate and the validation of reaction mechanism. For chemical reactions in condensed phase or enzymatic reactions, the computation of FE profile at \textit{ab initio} (\textit{ai}) quantum mechanical/molecular mechanics (QM/MM) level is still far too expensive. Although semiempirical (SE) method can be hundreds or thousands of times faster than the \textit{ai} methods, the accuracy of SE methods is often unsatisfactory, due to the approximations that have been adopted in these methods. In this work, we proposed a new method termed MBAR+wTP, in which the \textit{ai} QM/MM free energy profile is computed by a weighted thermodynamic perturbation (TP) correction to the SE profile generated by the multistate Bennett acceptance ratio (MBAR) analysis of the trajectories from umbrella samplings (US). The weight factors used in the TP calculations are a byproduct of the MBAR analysis in the post-processing of the US trajectories, which are often discarded after the free energy calculations. The raw \textit{ai} QM/MM free energy profile is then smoothed using Gaussian process regression, in which the noise of each datum is set to be inversely proportional to the exponential of the reweighting entropy. The results show that this approach can enhance the efficiency of \textit{ai} FE profile calculations by several orders of magnitude with only a slight loss of accuracy. This method can significantly enhance the applicability of \textit{ai} QM/MM methods in the studies of chemical reactions in condensed phase and enzymatic reactions.
\end{abstract}
\clearpage
  
\section{\label{sec:introduction}Introduction}	
Free energy profile (FE profile), defined as the free energy variation along the reaction coordinate(s) (RC), is now widely used to calculate the reaction rate and to examine the reaction mechanism.\cite{HuARPC2008} Based on the transition state theory (TST), the reaction rate is determined by the free energy barrier and the curvature of the free energy profile near the transition state.\cite{LaidlerJPC1983,TruhlarJPC1996} Due to the exponential decrease of the population with respect to the free energy, a brute force simulation is always insufficient for the states aways from the reactant and the product. Umbrella sampling (US) is now routinely employed to enhance the sampling of the system along the RC by restraining the RC around a series of values using biasing potentials. The trajectories are then analyzed using the multistate Bennett acceptance ratio (MBAR)\cite{ShirtsJCP2008,ShirtsarXiv2017} or its binned variant, the weighted histogram analysis method (WHAM)\cite{FerrenbergPRL1989,SouailleCPC2001,GallicchioJPCB2005,ChoderaJCTC2007} method, to construct the free energy profile. Multiple windows are required in the US sampling, and the number of windows is determined by the curvature of the free energy profile. MBAR and WHAM are global algorithms, which assume that a global equilibrium of all the windows have been reached.\cite{MeyPRX2014} Besides, the samples must be as statistically independent as possible. In order to obtain a statistically sound result, a long time simulation for each window is needed. \lpfrev{Another interesting method for data-processing that has emerged recently is the variational free energy profile (vFEP) method developed by Lee et al.\cite{LeeJCTC2013,LeeJCTC2014} In this method, free energy profile is parametrically constructed via a maximum-likelihood analysis of the fitting functions for regression. It may out-perform WHAM and MBAR when data are scarce and overlaps between neighboring windows are insufficient. Similar ideas were also reported in other works.\cite{MaraglianoJCP2008,MengJCTC2015}}
    
For the studies of chemical reactions in condensed phase, classical force fields, which lack a finite bond-dissociation energy, are in general not applicable. On the other end of the methodological spectrum, full quantum mechanical methods are not feasible either due to the extremely poor computational scaling with the system size. The hybrid quantum mechanical/molecular mechanical method (QM/MM)\cite{WarshelJMB1976} inherits the advantages from both sides, which makes it an appealing method and is nowadays routinely employed. However, the convergence of the FE profiles for chemical and enzymatic reactions in condensed phase is very slow, which requires exhaustive sampling in the high-dimensional phase space, and makes the computation of the FE profile at the \textit{ab initio} (\textit{ai}) QM/MM levels\cite{ZhangJCP1999,ZhangJCP2000,CuiJACS2001} notoriously time consuming, if feasible at all. Therefore, a great amount of QM/MM simulations employed semiempirical (SE) methods such as PM3\cite{StewartJCC1989}, AM1\cite{DewarJACS1985}, MNDO\cite{DewarJACS1977}, PM6\cite{StewartJMM2007}, and density functional theory-based tight-binding (DFTB)\cite{PorezagPRB1995,SeabraJPCA2007} as well as its self-consistent-charge variant (SCC-DFTB)\cite{ElstnerPRB1998}. Nevertheless, the results from QM/MM simulations employing an SE Hamiltonian are often of poor accuracy in terms of the reaction free energy and free energy barrier\cite{ChristensenJCP2017}. Consequently, developing an efficient method for the computation of state free energy and free energy profile at \textit{ai} QM/MM level remains one of the major challenges in modern computational chemistry.
  
In a pioneer work, Gao applied the dual-Hamiltonian method, aka the reference-potential method, in a study of the hydration free energy.\cite{GaoJPC1992,Gaoscience1992} Muller et al. and Bentzien et al. proposed another dual-Hamiltonian approach to calculate the FE profile for a chemical reaction\cite{MullerJPC1995,BentzienJPCB1998}, which was later named the paradynamics (PD) method.\cite{PlotnikovJPCB2011,PlotnikovJPCB2012,LameiraJPCB2016} In this method, an initial simulation is carried out utilizing empirical valence bond (EVB)\cite{WarshelJACS1980,Warshel1991,BilleterJCP2001} as a reference-potential, which is reasonably close to the real potential, allowing one to use a perturbation rectification in a subsequent step to obtain the free energy profile at the \textit{ai} level. However, constructing the EVB potential energy surface is nontrivial, and the simulations at high-level Hamiltonian are not completely avoided. Rod and Ryde computed the free energy barrier for a methyl transfer reaction catalyzed by the enzyme catechol O-methyltransferase using a dual Hamiltonian approach, in which the free energy profile sampled at a molecular mechanical level was corrected to the level of density functional theory using thermodynamic perturbation (TP).\cite{RodPRL2005} However, in their study, the QM degrees-of-freedom were not sampled, which inevitably led to overestimation of the free energy. Later on, Heimdal and Ryde studied the convergence of MM to QM/MM perturbation in the reference-potential method, and noticed that the perturbations typically have convergence problems, and a thoroughly parameterized force field or SQM methods are often required.\cite{HeimdalPCCP2012} However, the perturbation from the low-level Hamiltonian to the high-level Hamiltonian in this work was carried out using the traditional thermodynamic perturbation with equal weight assigned to the snapshots from biased simulations.
Beierlein et al applied this idea in the calculation of free energy in protein-ligand association.\cite{BeierleinJPCB2011} Polyak et al designed a dual Hamiltonian free energy perturbation (DH-FEP) method to calculate the free energy profiles of a model system and chorismate mutase.\cite{PolyakJCP2013} However, their method was not rigorous, because in the TP calculation the configurations were sampled on a low-level Hamiltonian but the energy difference between two adjacent windows were calculated on a high-level Hamiltonian. In addition, the constrained dynamics always overestimates the free energy difference in TP calculations. \lpfrev{More importantly, the configurations of the target RC (the unsampled state $\xi_{i+1}$ in Eq.~10 of ref.~\citenum{PolyakJCP2013}) were not unambiguous when being taken into the TP calculations.} Therefore, the calculation with two-dimensional RCs worked better than the one with a one-dimensional RC, as they have observed in their study. K\"{o}nig et al. combined energy reweighting with Bennett acceptance ratio and developed a new method called non-Boltzmann Bennett acceptance ratio (NBB). However, this method does not yield a minimum variance for the results.\cite{NBB,NBBJCTC} Hudson et al incorporated energy reweighting into the chain-of-replicas method for the computation of pathway free energy.\cite{HudsonBBA2015} Because the chain-of-replicas method is a global simulation method instead of a pure post-processing method, the implementation of the reweighting is not trivial. Hudson et al also used the non-equilibrium simulation method to correct the free energy profile of a low-level Hamiltonian to a high-level one.\cite{HudsonJPCL2015} However, this method requires expensive energy/force evaluation on the high-level Hamiltonian for every step of the simulation, which is not necessarily the preferred method in the first place for such studies.\cite{WangJCIM2017}
  
In our previous work\cite{JiaJCTC2016}, the dual-Hamiltonian approach was used to calculate the solvation free energies of the molecules in the SAMPL4 via an alchemical decoupling process. A classical force field was chosen as the reference-potential, on which level the solvation free energies were estimated using BAR or its multistate variant MBAR. Then a direct TP calculation was performed to obtain the MM-to-QM/MM correction. It has been shown that this (M)BAR+TP method yielded minimum variance for the free energy. Since this (M)BAR+TP method can be applied in the calculation of free energy difference between two thermodynamic states, quite naturally it can also be applied to the calculation of FE profile, because free energy difference between two thermodynamic states is calculated as a generalized FE profile in an extended phase space with one dimension of the coupling parameter appended to the normal phase space. Dybeck incorporated the dual Hamiltonian method with MBAR in the calculation of free energy difference between two states, which can be put forward to the calculations of FE profile.\cite{DybeckJCTC2016} Although this method yields the globally minimal variance for the free energies, it is an $O(N^2)$ algorithm, where $N$ is the number of the intermediate states.
 
Another class of approaches that has become a new boost in computational chemistry is machine learning (ML) or deep learning (DL).\cite{RuppIJQC2015,GohJCC2017} Based on some training using neural network, an energy correction term in a high-dimensional space (probably a complete phase space or a subspace of it) can be learned. Consequently, potential energy and/or force on the high-level Hamiltonian can be accurately estimated for the structural ensemble generated at the low-level Hamiltonian.\cite{HandleyJPCA2010,CaccinIJQC2015,RamakrishnanJCTC2015,BotuJPCC2017,ShenJCTC2016,WuJCP2017,DralJCP2017,LiJCTC2017} However, for most cases the learned energy/force correction for a chemical reaction is system-specific and short of transferability to other reactions, and the training itself is expensive. Even before the introduction of ML into this field, parameter tuning to high-level \textit{ai} accuracy for the SE methods and force fields to conduct the simulation has been performed by many groups.\cite{Gonzalez-LafontJPC1991,NamJCTC2007,DoronJCTC2011,ZhouJCTC2014,MaurerJCTC2007,WuJCTC2011,KnightJCTC2010}
  
\begin{figure}
	\centering
	\begin{subfigure} [b] {0.65\textwidth}
		\centering
		\includegraphics[width=0.65\textwidth]{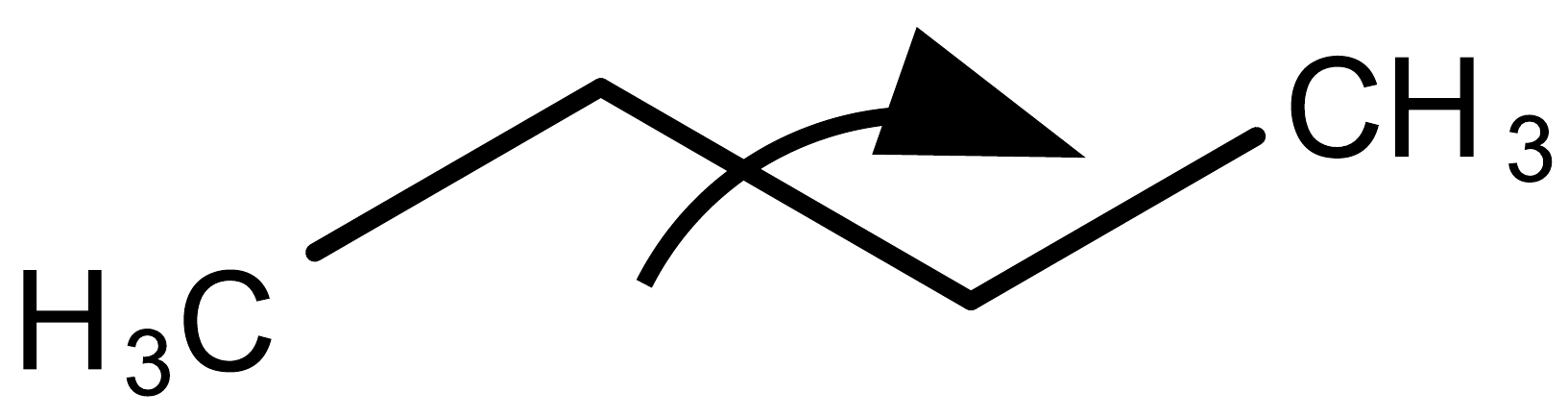}
		\caption{\label{fig:reaction1}}
	\end{subfigure}
	\centering
	\begin{subfigure} [b] {0.8\textwidth}
		\centering
		\includegraphics[width=0.8\textwidth]{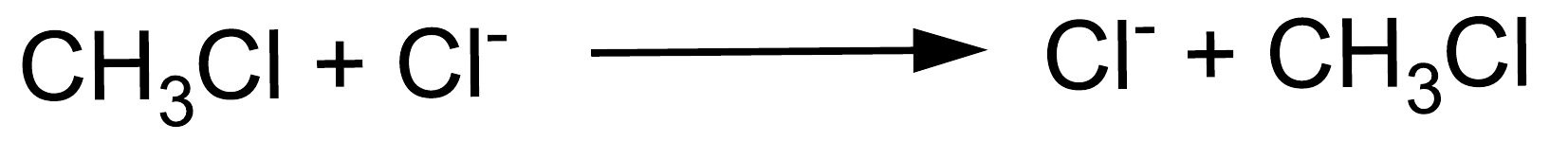}
		\caption{\label{fig:reaction2}}
	\end{subfigure}
	\centering
	\begin{subfigure} [b] {0.8\textwidth}
		\centering
		\includegraphics[width=0.8\textwidth]{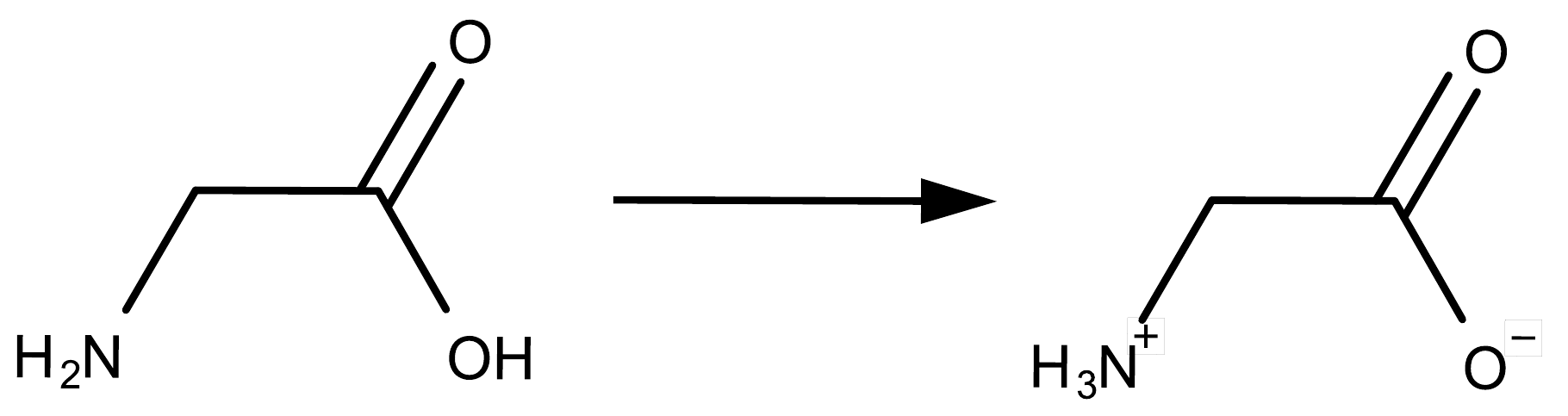}
		\caption{\label{fig:reaction3}}
	\end{subfigure} 
	\centering
	\begin{subfigure} [b] {0.8\textwidth}
		\centering
		\includegraphics[width=0.8\textwidth]{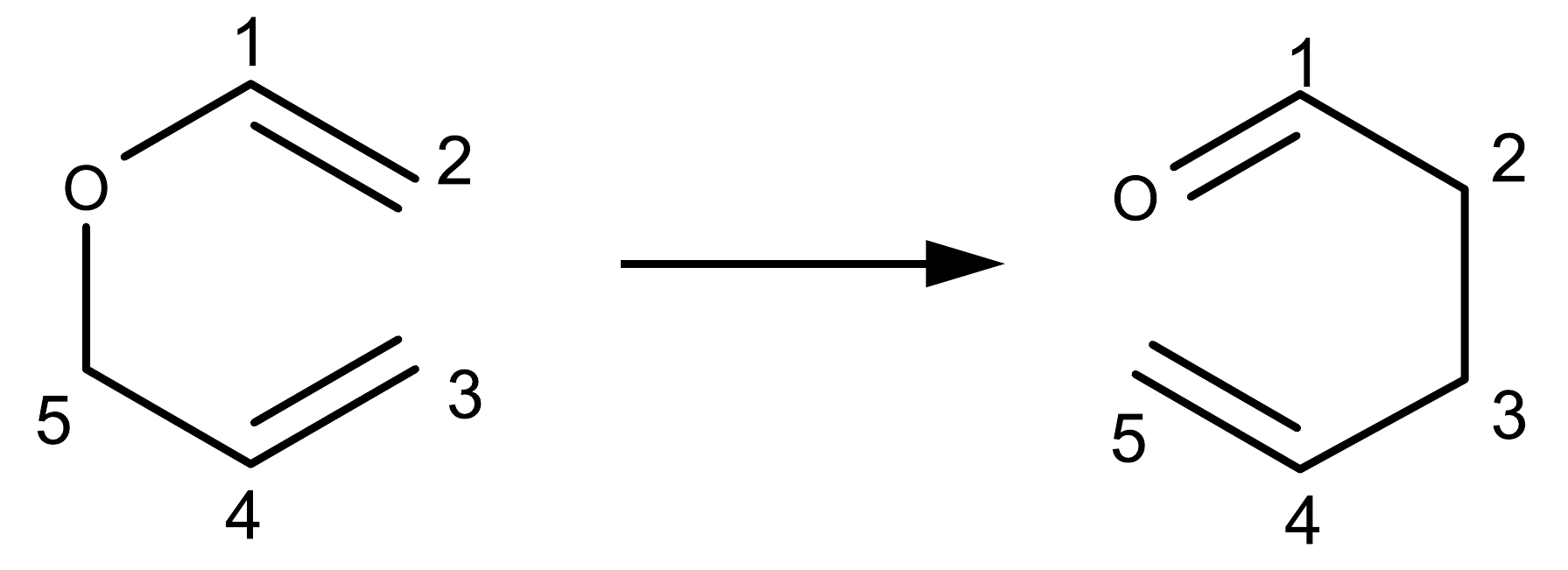}
		\caption{\label{fig:reaction4}}
	\end{subfigure}
	\caption{\label{fig:reactions} (a) The main chain dihedral rotation of a butane molecule, for which the reaction coordinate was chosen as $\eta = \phi\{C1-C2-C3-C4\}$. (b) The $\mathrm{S_{N} 2}$ reaction of \ce{CH3Cl + Cl- -> Cl- + CH3Cl}, for which the reaction coordinate was chosen as $\eta = d_{CCl1} - d_{CCl2}$. (c) The intramolecular proton transfer reaction of glycine from the neutral form (left) to the zwitterion form (right), for which the reaction coordinate was chosen as $\eta = d_{OH} - d_{NH}$. (d) The aliphatic Claisen rearrangement reaction of allyl vinyl ether (left) to 4-pentenal (right), for which reaction coordinate was chosen as $\eta = d_{OC5} - d_{C2C3}$.}
\end{figure}

In this work, we extended the idea in MBAR+TP to the calculations of the FE profiles of one quasi-chemical reaction and three chemical reactions in aqueous solution as shown in Fig.~\ref{fig:reactions}. Most importantly, we show that TP calculations should not be carried out in the traditional way, in which equal weights are assigned to the configurations. Instead, each configuration is assigned a weight factor from the MBAR analysis. The reactions studied include (a) main chain dihedral rotation of a butane molecule, (b) $\mathrm{S_{N} 2}$ reaction of \ce{CH3Cl + Cl- -> Cl- + CH3Cl}, (c) glycine intramolecular proton transfer reaction from the neutral form to the zwitterion form, and (d) aliphatic Claisen rearrangement reaction of allyl vinyl ether to 4-pentenal. We show that the FE profile of high-level Hamiltonian can be obtained in this way with configuration sampling carried out only using the low-level Hamiltonian, and only one hundredth of the configurations were used for the high-level energy calculations. As a consequence, the formidably expensive sampling using the high-level Hamiltonian was avoided, thus enhancing the efficiency by several orders of magnitude. 
Inevitably, the raw free energy profile computed from this weighted TP are contaminated by statistical noise. Gaussian process regression was applied using the statistical noise, as well as the profile itself, as the input. A smoothed profile and its uncertainty can thus be obtained.

\section{\label{sec:method:MBARandTP}Method}
\subsection{\label{sec:USandMBAR}Umbrella Sampling and Multistate Bennett Acceptance Ratio}
For reactions in condensed phase, the reactant and the product states are often separated by a free energy barrier hindering the transition between these two states. In order to characterize this free energy barrier, a reaction path defined by reaction coordinates (RC) $\eta(\textbf{x})$ in a low dimension should be carefully chosen, and the free energy profile along the reaction path from the reactant to the product is proportional to the logarithm of the probability of visit by the system. The reaction coordinate $\eta(\textbf{x})$ is usually a function of the collective atomic coordinates $\textbf{x}$. The reaction rate can be estimated by the ratio of the free energy barrier over $k_{B}T$, where $k_{B}$ is Boltzmann constant and $T$ is the temperature. When the reaction barrier is much greater than $k_{B}T$, which is often the case for chemical reactions under mild conditions, the probability of visit to the regions away from either the reactant or the product is rare, especially the visit to the transition state, and computational cost for a brute force simulation is far beyond what we can afford with modern computers. Umbrella sampling (US) method\cite{TorrieJComputP1977} is a solution to this situation, and it is routinely used nowadays. US is a stratified approach, and each stratum samples a limited range of phase space along the RC by augmenting the original potential energy surface $U_{0}(\textbf{x})$ with a restraining bias $W_{i}(\eta)$. A harmonic potential is frequently used as the bias $W_{i}(\eta)=\frac{1}{2}k_{i}(\eta-\eta_{i})^2$, where $\eta_{i}$ and $k_{i}$ are the target value of RC and the strength of the restraint, respectively, for the $i$th biased window. \lpfrev{Both the strength of restraint $k_{i}$ value and the number of US windows can affect the magnitude of overlap between adjacent windows, which is critical to the quantity of free energy calculations. In this work, $k_{i}$ values were optimized for each window based on short trial simulations, and overlap between neighboring windows should be larger than 0.03 using an assessment proposed by Klimovich et al. (See SI for details).} The trajectories from a set of biased window simulations indexed by $i$ ($i=1,2,\dots,S$) with the potential energy surfaces $U_{i}^{(b)}(\eta) = U_{0}(\textbf{x}) + W_{i}(\eta(\textbf{x}))$ are then post-processed using the Weighted Histogram Analysis Method (WHAM)\cite{FerrenbergPRL1989,SouailleCPC2001,GallicchioJPCB2005,ChoderaJCTC2007} or its ``binless'' variant, the Multistate Bennett Acceptance Ratio (MBAR)\cite{ShirtsJCP2008,ShirtsarXiv2017}, globally to obtain the unbiased thermodynamic properties on the original potential energy surface $U_0$. Both methods have been well-documented in the literature \cite{FerrenbergPRL1989,SouailleCPC2001,GallicchioJPCB2005,ChoderaJCTC2007,ShirtsJCP2008,ShirtsarXiv2017}. 
Here, we will present a brief explanation of the MBAR method for the analysis of US simulations. 
    
According to the MBAR method\cite{ShirtsJCP2008,ShirtsarXiv2017}, the free energy $f_{t}^{(b)}$ of the $t$th biased simulation window can be obtained by
\begin{align}
e^{-\beta f_{t}^{(b)}}=&\sum\limits_{i=1}^{S}\sum\limits_{l=1}^{N_{i}}\frac{e^{-\beta U_{t}^{(b)}(\textbf{x}_{i,l})}}{\sum\limits_{k=1}^{S}N_{k}e^{-\beta\left[U_{k}^{(b)}(\textbf{x}_{i,l})-f_{k}^{(b)}\right]}} \notag\\
=&\sum\limits_{i=1}^{S}\sum\limits_{l=1}^{N_{i}}\frac{e^{-\beta \left[U_{0}(\textbf{x}_{i,l})+W_{t}(\textbf{x}_{i,l})\right]}}{\sum\limits_{k=1}^{S}N_{k}e^{-\beta\left[U_{0}(\textbf{x}_{i,l})+W_{k}(\textbf{x}_{i,l})-f_{k}^{(b)}\right]}} \notag\\
=&\sum\limits_{i=1}^{S}\sum\limits_{l=1}^{N_{i}}\frac{e^{-\beta \left[W_{t}(\textbf{x}_{i,l})\right]}}{\sum\limits_{k=1}^{S}N_{k}e^{-\beta\left[W_{k}(\textbf{x}_{i,l})-f_{k}^{(b)}\right]}},
\label{Eq:mbar}
\end{align}
where $\textbf{x}_{i,l}$ is the $l$th frame among $N_{i}$ configurations in the $i$th biased simulation. Since ${f_{t}^{(b)}}$ appear on both the left-hand side and the right-hand side of Eq.~\ref{Eq:mbar}, \lpfrev{these equations were solved in this work by a simple self-consistent iteration method using an in-house program, although some other methods are also available\cite{ShirtsJCP2008}.} Sorting the samples into bins are unnecessary. In other words, MBAR is equivalent to WHAM in the limit that bin widths shrunk to zero and can be derived without the need to invoke histograms.\cite{ShirtsJCP2008} It is worth emphasizing that $f_t^{(b)}$ is not necessarily among $\{f_k^{(b)}, k=1,\dots,K\}$. It can be the free energy of any state of the system with a potential energy function $U_{t}(\textbf{x})=U_0(\textbf{x})+W_t(\textbf{x})$. In particular, if $W_t(\textbf{x})=0$, $U_{t}(\textbf{x})$ equals to the unbiased potential energy $U_0(\textbf{x})$, and $f_t^{(b)}$ becomes the unbiased free energy $f_0$. 

The weight of the $l$th configuration in the $i$th biased simulation for any Hamiltonian $U_t$ is, akin to Eq.~9 in ref.~\citenum{ShirtsJCP2008},
\begin{align}
w_{t}(\textbf{x}_{i,l})=&\frac{e^{-\beta \left[U_{t}(\textbf{x}_{i,l})-f_{t}\right]}}{\sum\limits_{k=1}^{S}N_{k}e^{-\beta\left[U_{k}^{(b)}(\textbf{x}_{i,l})-f_{k}^{(b)}\right]}}.
\label{Eq:weight} 
\end{align}
This definition ensures that $\sum\limits_{i=1}^{S}\sum\limits_{l=1}^{N_{i}}w_{t}(\textbf{x}_{i,l})=1$ for all $t=1,2,\dots,S$ and $\sum\limits_{t=1}^{S}N_{t}w_{t}(\textbf{x}_{i,l})=1$ for all $\textbf{x}_{i,l}, $ where $ i=1,2,\dots,S; l=1,2,\dots,N_{i}$. 
The unbiased probability $\rho_{0}(\eta)$ can be calculated now by 
\begin{align}
\rho_{0}(\eta) = & \sum\limits_{i=1}^{S}\sum\limits_{l=1}^{N_{i}} \frac{\delta {\left(\textbf{x}_{\eta(i,l)}-\textbf{x}_{i,l}\right)}}{\sum\limits_{k=1}^{S}N_{k}e^{-\beta\left[W_{k}(\textbf{x}_{i,l})-f_{k}^{(b)}\right]}},
\label{Eq:unbiasx2}
\end{align}
and the free energy profile is calculated as
\begin{align}
F(\eta)=& -\beta^{-1}\ln{\rho_{0}(\eta)} \notag\\
       =& -\beta^{-1}\ln{\sum\limits_{i=1}^{S}\sum\limits_{l=1}^{N_{i}} \frac{\delta {\left(\textbf{x}_{\eta(i,l)}-\textbf{x}_{i,l}\right)}}{\sum\limits_{k=1}^{S}N_{k}e^{-\beta\left[W_{k}(\textbf{x}_{i,l})-f_{k}^{(b)}\right]}}}.
\label{Eq:MBARFEProfile}
\end{align}    

The covariance matrix can be obtained from Eq.~D8 in Ref.~\citenum{ShirtsJCP2008} by
\begin{align}
\boldsymbol{\Theta}
=&{\left[\left({\textbf{W}}^{T}\textbf{W}\right)^{-1} - \textbf{N} + \textbf{1}_{S+M}\textbf{1}_{S+M}^{T}/N\right]}^{-1},
\label{Eq:MBARvar1}
\end{align}
where $\textbf{N}$ is a $(S+M) \times (S+M) $ matrix, $\textbf{N}=diag(N_{1},N_{2},\dots,N_{S},0_{1},0_{2},\dots,0_{M}),$ $M$ denotes the number of histogram bins and $N = \sum\limits_{i=1}^{S}\sum\limits_{l=1}^{N_{i}} 1$ is the total number of samples collected from all the biased simulations, and $\textbf{W}$ is the weight matrix with a dimension of $N \times (S+M) $. The $N \times S$ elements of $\textbf{W}$ can be obtained via
\begin{align}
w_{i}(\textbf{x}_{i,l})
=&\frac{e^{-\beta\left[W_{i}(\textbf{x}_{i,l})-f_{i}^{(b)}\right]}}{\sum\limits_{k=1}^{S}N_{k}e^{-\beta\left[W_{k}(\textbf{x}_{i,l})-f_{k}^{(b)}\right]}}.
\label{Eq:weight1} 
\end{align}
The $N \times M$ elements of $\textbf{W}$ can be obtained by
\begin{align}
w_{\eta}(\textbf{x}_{i,l})
=&\frac{\frac{\delta {\left(\textbf{x}_{\eta(i,l)}-\textbf{x}_{i,l}\right)}}{\sum\limits_{k=1}^{S}N_{k}e^{-\beta\left[W_{k}(\textbf{x}_{i,l})-f_{k}^{(b)}\right]} } } {\sum\limits_{i=1}^{S}\sum\limits_{l=1}^{N_{i}}{\frac{\delta {\left(\textbf{x}_{\eta(i,l)}-\textbf{x}_{i,l}\right)}}{\sum\limits_{k=1}^{S}N_{k}e^{-\beta\left[W_{k}(\textbf{x}_{i,l})-f_{k}^{(b)}\right]}}}}.
\label{Eq:weight2} 
\end{align}
The uncertainty in the estimated free energy difference can be computed as
\begin{align}
\delta^{2} \Delta F\left(\eta_{ij}\right) = \boldsymbol{\Theta}_{ii} - 2 \boldsymbol{\Theta}_{ij} + \boldsymbol{\Theta}_{jj}.
\end{align}

\subsection{\label{sec:TP}Weighted Thermodynamic Perturbation}
Thermodynamic Perturbation (TP), also known as Free Energy Perturbation, exponential average, and Zwanzig equation was developed by Zwanzig.\cite{ZwanzigJCP1954}. In TP, only one simulation (normally with the inexpensive Hamiltonian) is required and the free energy difference between the target Hamiltonian and the sampled Hamiltonian can be calculated with the sampled configurations. Using the binned configurations from the US simulations and their weights from the MBAR analysis, the FE profile at high level Hamiltonian can be calculated by TP. According to Eq.~\ref{Eq:weight}, the weight for the $l$th configuration in the $i$th biased simulation under the low-level Hamiltonian, which is the unbiased SE Hamiltonian in this study, is
\begin{align}
w_L(\textbf{x}_{i,l})=&\frac{e^{-\beta \left[U_L(\textbf{x}_{i,l})-f_L\right]}}{\sum\limits_{k=1}^{S}N_{k}e^{-\beta\left[U_{k}^{(b)}(\textbf{x}_{i,l})-f_{k}^{(b)}\right]}}\notag\\
=&\frac{e^{\beta f_L}}{\sum\limits_{k=1}^{S}N_{k}e^{-\beta\left[W_{k}(\textbf{x}_{i,l})-f_{k}^{(b)}\right]}},
\label{Eq:weightL} 
\end{align}
where $e^{\beta f_L}$ is constant for a given Hamiltonian, which can be canceled in normalization. In a ``traditional'' unbiased TP calculation, all samples would have equal weights, which is actually a special case of Eq.~\ref{Eq:weightL}. Suppose we have performed a series of brute-force unbiased simulations to sample the configurational phase covering the reactant, the product and in between. $W_k=0$ for all the snapshots, $f_k^{(b)}$ becomes the unbiased free energy, and the exponential terms in the numerator and denominator cancel each other. We find $w_L=1/\sum\limits_{k=1}^{S}N_{k}=1/N$, indicating that all the samples have equal weights in a ``traditional'' unbiased TP calculation. 
     
For a certain histogram bin, the free energy difference between the high-level and low-level Hamiltonians can be obtained via
\begin{align}
\Delta F (\eta) =& -\frac{1}{\beta} \ln \frac{\sum\limits_{i=1}^{S}\sum\limits_{l=1}^{N_{i}} w_{L}\left(\mathbf{x}_{\eta}\right)\delta\left(\eta_{(i,l)}-\eta\right)e^{-\beta\left[U_{H}\left(\mathbf{x}_{\eta}\right) - U_{L}\left(\mathbf{x}_{\eta}\right)\right]}}{\sum\limits_{i=1}^{S}\sum\limits_{l=1}^{N_{i}} w_{L}\left(\mathbf{x}_{\eta}\right)\delta\left(\eta_{(i,l)}-\eta\right)} \notag\\
=& -\frac{1}{\beta} \ln \frac{\sum\limits_{i=1}^{S}\sum\limits_{l=1}^{N_{i}} w_{L}\left(\mathbf{x}_{\eta_{(i,l)}}\right)e^{-\beta\left[U_{H}\left(\mathbf{x}_{\eta_{(i,l)}}\right) - U_{L}\left(\mathbf{x}_{\eta_{(i,l)}}\right)\right]} }{\sum\limits_{i=1}^{S}\sum\limits_{l=1}^{N_{i}} w_{L}\left(\mathbf{x}_{\eta_{(i,l)}}\right)},
\label{Eq:deltaF}
\end{align}
where the subscripts $H$ and $L$ denote the high-level and the low-level Hamiltonians, respectively, $\eta_{(i,l)}$ is the histogram bin that the $l$th configuration in the $i$th biased simulation belongs to, and $w_L(\mathbf{x}_{\eta})$ is the unbiased weight of the sample in bin $\eta$ under the low-level Hamiltonian. Here, the delta function singles out the samples inside bin $\eta$. It is worth emphasizing that , as a test of our methodology, we can also calculate the weights of the samples under the high-level Hamiltonian, and then compute the free energy difference between the low-level Hamiltonian and the high level Hamiltonian in a reverse way. Or we can calculate the weights of the samples under both the low-level and high-level Hamiltonians and then compute the free energy via BAR. All these calculations yield the same results as shown in SI. 

In order to obtain the uncertainty $\delta ^2 \Delta F(\eta)$, Eq.~\ref{Eq:deltaF} can be divided into two terms as
\begin{align}
\Delta F (\eta)
=& -\frac{1}{\beta} \ln{\sum\limits_{i=1}^{S}\sum\limits_{l=1}^{N_{i}} w_{L}\left(\textbf{x}_{\eta_{(i,l)}}\right)e^{-\beta\left[U_{H}\left(\textbf{x}_{\eta_{(i,l)}}\right) - U_{L}\left(\textbf{x}_{\eta_{(i,l)}}\right)\right]}} \notag\\
&+ \frac{1}{\beta} \ln{\sum\limits_{i=1}^{S}\sum\limits_{l=1}^{N_{i}} w_{L}\left(\textbf{x}_{\eta_{(i,l)}}\right)}.
\label{Eq:deltaF2}
\end{align}
The uncertainty $\delta ^2 \Delta F(\eta)$ can be computed by error propagation
\begin{equation}
\delta ^2 \Delta F(\eta) = \delta ^2_I + \delta ^2_{II}.
\label{Eq:var}
\end{equation}
Here, $\delta ^2_I$ and $\delta ^2_{II}$ correspond to the first and the second terms on the right-hand side of Eq.~\ref{Eq:deltaF2}, respectively.
\begin{equation}
\delta ^2_I = \frac{1}{\beta ^2}\cdot\frac{1}{\left(\sum\limits_{i=1}^{S}\sum\limits_{l=1}^{N_{i}} \delta\left(\eta_{(i,l)}-\eta\right)\right)^2}\cdot \frac{\sigma^{2}_{y^{1}_{i,l}}}{{\left\langle y^{1}_{i,l} \right\rangle}^2},
\label{Eq:var1}
\end{equation}
where $y^{1}_{i,l} = w_{L}(\textbf{x}_{\eta_{(i,l)}})e^{{-\beta\left[U_{H}\left(\textbf{x}_{\eta_{(i,l)}}\right) - U_{L}\left(\textbf{x}_{\eta_{(i,l)}}\right)\right]}}$, $\sum\limits_{i=1}^{S}\sum\limits_{l=1}^{N_{i}} \delta\left(\eta_{(i,l)}-\eta\right)$ is the number of configurations falling into bin $\eta$ among the samples from all the biased simulations.
\begin{equation}
\delta ^2_{II} = \frac{1}{\beta ^2}\cdot\frac{1}{\left(\sum\limits_{i=1}^{S}\sum\limits_{l=1}^{N_{i}} \delta\left(\eta_{(i,l)}-\eta\right)\right)^2}\cdot \frac{\sigma^{2}_{y^{2}_{i,l}}}{{\left\langle y^{2}_{i,l} \right\rangle}^2},
\label{Eq:var2}
\end{equation}
where $y^{2}_{i,l} = w_{i}\left(\textbf{x}_{\eta_{(i,l)}}\right)$, $\sum\limits_{i=1}^{S}\sum\limits_{l=1}^{N_{i}} \delta\left(\eta_{(i,l)}-\eta\right)$ is the number of configurations pertaining to histogram bin $\eta$ among all the samples.
     
In this work, the US samplings were performed at the low-level Hamiltonian, of which the FE profile $F_{L}(\eta)$ can be obtained. Then, the FE profile of the high-level Hamiltonian can be calculated by
\begin{equation}
F_{H}(\eta)=F_{L}(\eta) + \Delta F (\eta).
\label{Eq:FH}
\end{equation}
The corresponding uncertainty of $F_{H}(\eta)$ can be computed by error propagation
\begin{equation}
\delta^{2} F_{H}(\eta) = \delta^{2} F_{L}(\eta) + \delta^{2} \Delta F(\eta).
\end{equation}
In the following, we will use direct FE profile to refer to the FE profile obtained from the US samplings at the corresponding Hamiltonian, and use indirect FE profile to refer to the one obtained from TP correction over the direct FE profile from a different Hamiltonian. 
     
In order to characterize the reliability of the TP calculation, ``reweighting entropy'' is introduced in our previous work\cite{WangJCIM2017}, which is defined as
\begin{equation}
\mathcal{S}(\eta) = -\frac{1}{\ln\left( \sum\limits_{i=1}^{S}\sum\limits_{l=1}^{N_{i}}\delta (\eta_{(i,l)}-\eta)\right)}\sum\limits_{i=1}^{S}\sum\limits_{l=1}^{N_{i}}\mathcal{P}(\eta_{(i,l)})\ln{\mathcal{P}(\eta_{(i,l)})},
\label{Eq:RWentropy}
\end{equation}
where  
\begin{align}
\mathcal{P}\left(\eta_{(i,l)}\right)
%=&\frac{\overline{w}_{L}\left(\textbf{x}_{\eta_{(i,l)}}\right)e^{-\beta\left[U_{H}\left(\textbf{x}_{\eta_{(i,l)}}\right) - U_{L}\left(\textbf{x}_{\eta_{(i,l)}}\right)\right]}}{\sum\limits_{i=1}^{S}\sum\limits_{l=1}^{N_{i}} \overline{w}_{L}\left(\textbf{x}_{\eta_{(i,l)}}\right)e^{-\beta\left[U_{H}\left(\textbf{x}_{\eta_{(i,l)}}\right) - U_{L}\left(\textbf{x}_{\eta_{(i,l)}}\right)\right]}}\notag\\
=&\frac{w_{L}\left(\textbf{x}_{\eta_{(i,l)}}\right)e^{-\beta\left[U_{H}\left(\textbf{x}_{\eta_{(i,l)}}\right) - U_{L}\left(\textbf{x}_{\eta_{(i,l)}}\right)\right]}}{\sum\limits_{i=1}^{S}\sum\limits_{l=1}^{N_{i}} w_{L}\left(\textbf{x}_{\eta_{(i,l)}}\right)e^{-\beta\left[U_{H}\left(\textbf{x}_{\eta_{(i,l)}}\right) - U_{L}\left(\textbf{x}_{\eta_{(i,l)}}\right)\right]}}.
\label{Eq:Wmax}
\end{align}
%and
%\begin{align}
%\overline{w}_{L}\left(\textbf{x}_{\eta_{(i,l)}}\right)=&\frac{w_{L}\left(\textbf{x}_{\eta_{(i,l)}}\right)}{\sum\limits_{i=1}^{S}\sum\limits_{l=1}^{N_{i}} %{w_{L}\left(\textbf{x}_{\eta_{(i,l)}}\right)}}.
%\label{Eq:wLbar}
%\end{align}
Here, $\sum\limits_{i=1}^{S}\sum\limits_{l=1}^{N_{i}} \delta \left(\eta_{(i,l)}-\eta\right)$ represents the number of configurations falling into bin $\eta$ among the samples from all the biased simulations. The larger the reweighting entropy is, the more reliable the TP calculation can be. $\mathcal{S}$ has a maximum value of 1.0, where all the samples have equal weights. The maximum of $\mathcal{P}(\eta_{(i,l)})$ among $\sum\limits_{i=1}^{S}\sum\limits_{l=1}^{N_{i}} \delta \left(\eta_{(i,l)}-\eta\right)$ frames is called ``maximal weight'' $\mathcal{P}_{max}$. Generally speaking, the smaller the maximal weight is, the more reliable the TP calculation can be. It is well known that TP requires significant overlap in phase space between the sampled (PM3 or PM6 in this work) Hamiltonian and the target (PM6 or B3LYP in this work) Hamiltonian.\cite{LuFECbook} 
%Otherwise, the TP calculation is unreliable. 
%It is worth emphasizing that $\overline{w}_{L}\left(\textbf{x}_{\eta_{(i,l)}}\right)$ changes the relative weight of each sample among the samples from all the bins, but not the relative weight for samples in the same bin.
Recently, we noticed that a similar quantity has been also suggested by Hansen et al.\cite{HunenbergerJCTC2010}

\subsection{\label{sec:regression}Gaussian process regression for curve fitting}
Gaussian Processes (GP) are generic supervised learning methods designed to solve regression and probabilistic classification problems.\cite{GPRbook} It is an attractive approach because it is nearly model-free. Gaussian process regression (GPR) has been used in free energy surface reconstruction.\cite{US_GPR,FES_GPR}
In this work, Gaussian process regression was utilized to fit the free energy profiles after the weighted TP correction, which is always contaminated by statistical noise. Given a set of observations $\{\Delta F_{1},\Delta F_{2},...,\Delta F_{n}\}$, it can be imagined as a single sample from a Gaussian distribution with $n$ variates. A Gaussian process is a collection of random variables, any finite number of which have a joint Gaussian distribution. A Gaussian process $f(\eta)$ can be completely specified by its mean $m(\eta)$ and covariance function $k(\eta,\eta^{'})$, expressed as
\begin{align}
	m(\eta)=E\left[f(\eta)\right]
\end{align}
and
\begin{align}
	k(\eta,\eta^{'})=E\left[\Big(f(\eta)-m(\eta)\Big)\left(f(\eta^{'})-m(\eta^{'})\right)\right],
\end{align}
where $E \left [\cdot \right ]$ denotes the expectation operation and the covariance function $k(\eta,\eta^{'})$ relates one observation to another. Since the observations are noisy, each observation $\Delta F$ is related to an underlying function $f(\eta)$ through a Gaussian noise model
\begin{align}
	\Delta F = f(\eta)+\mathcal{N}(0,\sigma^{2}_{n} ).
\end{align}
By considering also the noise, the covariance function $k$ was defined using the squared exponential as
\begin{align}
	k(\eta,\eta^{'})=\sigma^{2}_{f}\exp\left[\frac{-(\eta-\eta^{'})^{2}}{2l^{2}}\right] + \alpha \sigma^{2}_{n}\delta(\eta,\eta^{'}),
\end{align}
where $l$ is the length-scale and $\sigma^{2}_{f}$ is the signal variance, $\delta(\eta,\eta^{'})$ is the Kronecker delta function and $\sigma^{2}_{n}$ is the noise variance, which was set to the reciprocal of the exponential of the reweighting entropy value ($e^{-\mathcal{S}}$) corresponding to each observation in this work. 
%It is noted that the covariance between the outputs (observations) can be written as a function of the inputs. 
The free parameters $\{l,\sigma_{f},\alpha\}$ are the ``hyperparameters'' that are optimized to maximize the likelihood of the observations.
%The parameter $\alpha$ can be thought as a scale factor, which can scale the effect of noise and has a different value in the different weighted TP correction process. 
%The Gaussian process can be expressed as
%\begin{align}
%	f(x) \sim \mathcal{GP}\left(m(x),k(x,x^{'})\right)
%\end{align}
%The goal of regression is to search for $f(x)$. 
%This is a reasonable choice, because the larger the reweighting entropy is, the more reliable the observation is, equivalently, the smaller the noise for the observation is.
The Gaussian process regression was carried out using the scikit-learn package\cite{scikit-learn}.

\subsection{\label{sec:method:simulation}Molecular Dynamics Simulations}
One quasi-chemical reaction and three chemical reactions in aqueous solution shown in Fig.~\ref{fig:reactions} are studied, including main chain dihedral rotation of a butane molecule, an $\mathrm{S_{N} 2}$ reaction of \ce{CH3Cl + Cl- -> Cl- + CH3Cl}, intramolecular proton transfer in glycine from the neutral form to the zwitterion form, and an aliphatic Claisen rearrangement reaction from allyl vinyl ether to 4-pentenal.
     
For the main chain dihedral rotation of a butane molecule, the entire molecule was defined as the semiempirical QM (SQM) or QM region. 
For the $\mathrm{S_{N} 2}$ reaction, the complex of \ce{CH3Cl} and \ce{Cl-} was defined as the semiempirical QM (SQM) or QM region. 
For the glycine intramolecular proton transfer reaction, the glycine molecule was defined as the SQM or QM region. 
For the Claisen rearrangement reaction, the solute was defined as the SQM or QM region. 
A TIP3P water sphere with a radius of 25 {\AA} was added to the solute centering on the heavy atom closest to the center-of-mass of the QM regions and was restrained by a soft half-harmonic potential with a force constant of 10 kcal/mol/$\mathrm{{\AA}^2}$ to avoid evaporation. Thiel et al. have justified the widespread use of a droplet model in QM/MM studies of reactions in solution and in enzymes\cite{VasilevskayaJCTC2016}. \lpfrev{There were 2016 water molecules for the butane molecule system, 2014 water molecules for the $\mathrm{S_{N} 2}$ reaction, 2017 water molecules for the glycine intramolecular proton transfer reaction and 2009 water molecules for the Claisen rearrangement reaction.} The nonbonded interactions were fully counted without any truncations. The van der Waals (vdW) interactions were described with the general AMBER force field\cite{WangJCC2004}. PM3 and PM6 were used as the low-level Hamiltonians in the umbrella samplings, and the high-level QM Hamiltonian was chosen as B3LYP/6-31G(d). \lpfrev{In order to measure the reliability of this indirect method, umbrella samplings at B3LYP level were also performed.} We first compared the direct FE profile at PM6 with the indirect one at the same level reweighted from PM3 level. Then, the direct free energy profile at B3LYP/6-31G(d) level was computed to validate the indirect FE profile at the same level reweighted from the SE levels. 
          
For the main chain dihedral rotation of a butane molecule, the main chain dihedral $\eta = \phi\{\mathrm{C1-C2-C3-C4}\}$ was chosen as the reaction coordinate. Umbrella samplings with 61 windows centering on $\eta$ ranged from 0 to 180 degrees were performed. \lpfrev{The force constant $k_{i}$ of the bias restraint ranged from 100 kcal/mol/$\mathrm{{deg}^2}$ to 150 kcal/mol/$\mathrm{{deg}^2}$ in PM3, PM6 and B3LYP simulations for this system.} 
The reaction coordinate for the $\mathrm{S_{N} 2}$ reaction was defined as $\eta = d_{CCl1} - d_{CCl2}$, where $d_{CCl1}$ and $d_{CCl2}$ are the bond distances. Umbrella samplings with 73 windows centering on $\eta$ ranged from -2.5 to 2.5 $\mathrm{\AA}$ were performed. \lpfrev{The force constant $k_{i}$ of the bias restraint ranged from 100 kcal/mol/$\mathrm{{\AA}^2}$ to 900 kcal/mol/$\mathrm{{\AA}^2}$ in PM3 simulations and from 100 kcal/mol/$\mathrm{{\AA}^2}$ to 800 kcal/mol/$\mathrm{{\AA}^2}$ in PM6 and B3LYP simulations for this system.}
The reaction coordinate for the glycine intramolecular proton transfer reaction was defined as $\eta = d_{OH} - d_{NH}$, where $H$ is the hydrogen atom to be transferred. Umbrella samplings with 61 windows under PM3 level and with 51 windows under PM6 and B3LYP levels centering from on $\eta$ ranged from -1.5 to 1.5 $\mathrm{\AA}$ were applied. \lpfrev{The force constant $k_{i}$ of the bias restraint ranged from 100 kcal/mol/$\mathrm{{\AA}^2}$ to 1350 kcal/mol/$\mathrm{{\AA}^2}$ in PM3 simulations and from 100 kcal/mol/$\mathrm{{\AA}^2}$ to 900 kcal/mol/$\mathrm{{\AA}^2}$ in PM6 and B3LYP simulations for this system.}
The reaction coordinate for the Claisen rearrangement reaction was defined as $\eta = d_{OC5} - d_{C2C3}$. Umbrella samplings with 95 windows centering on $\eta$ ranged from -2.2 to 1.7 $\mathrm{\AA}$ were applied. \lpfrev{The force constant $k_{i}$ of the bias restraint ranged from 100 kcal/mol/$\mathrm{{\AA}^2}$ to 1600 kcal/mol/$\mathrm{{\AA}^2}$ in PM3 simulations and from 100 kcal/mol/$\mathrm{{\AA}^2}$ to 1400 kcal/mol/$\mathrm{{\AA}^2}$ in PM6 and B3LYP simulations for this system.}
For each US window simulation, the system was optimized in energy for 500 steps using the steepest decent optimization method followed by 500 steps of the conjugate gradient method with the respective Hamiltonian. For SE Hamiltonian, the system was heated up to 300 K in 50 ps and was equilibrated for 100 ps. A 1-ns production simulation was conducted for each window for free energy analysis. However, for B3LYP Hamiltonian the system was heated up to 300 K in 10 ps and was equilibrated for 10 ps. The production simulation has a length of 100-ps for each window. The integration time step was set to 1 fs, except for that in the PM3 and B3LYP simulations of the glycine intramolecular proton transfer reaction, which was set to 0.5 fs. The configurations were saved every 0.1 ps in the PM3 and PM6 simulations, and were saved every 0.05 ps in the B3LYP simulations. The temperature was regulated at 300 K with the Andersen temperature coupling scheme\cite{TariqJCP1983}. 
All the simulations were performed by the AmberTools 16 program package\cite{Amber16}, and the QM/MM calculations were carried out by interfacing with Q-Chem 4.3 package\cite{Q-Chem4}.
       
\section{\label{sec:result}Results}
\subsection{\label{sec:result:BT}Main chain dihedral rotation of a butane molecule}
\begin{figure}
	\begin{subfigure}{.5\textwidth}
		\centering
		\includegraphics[width=1.0\linewidth]{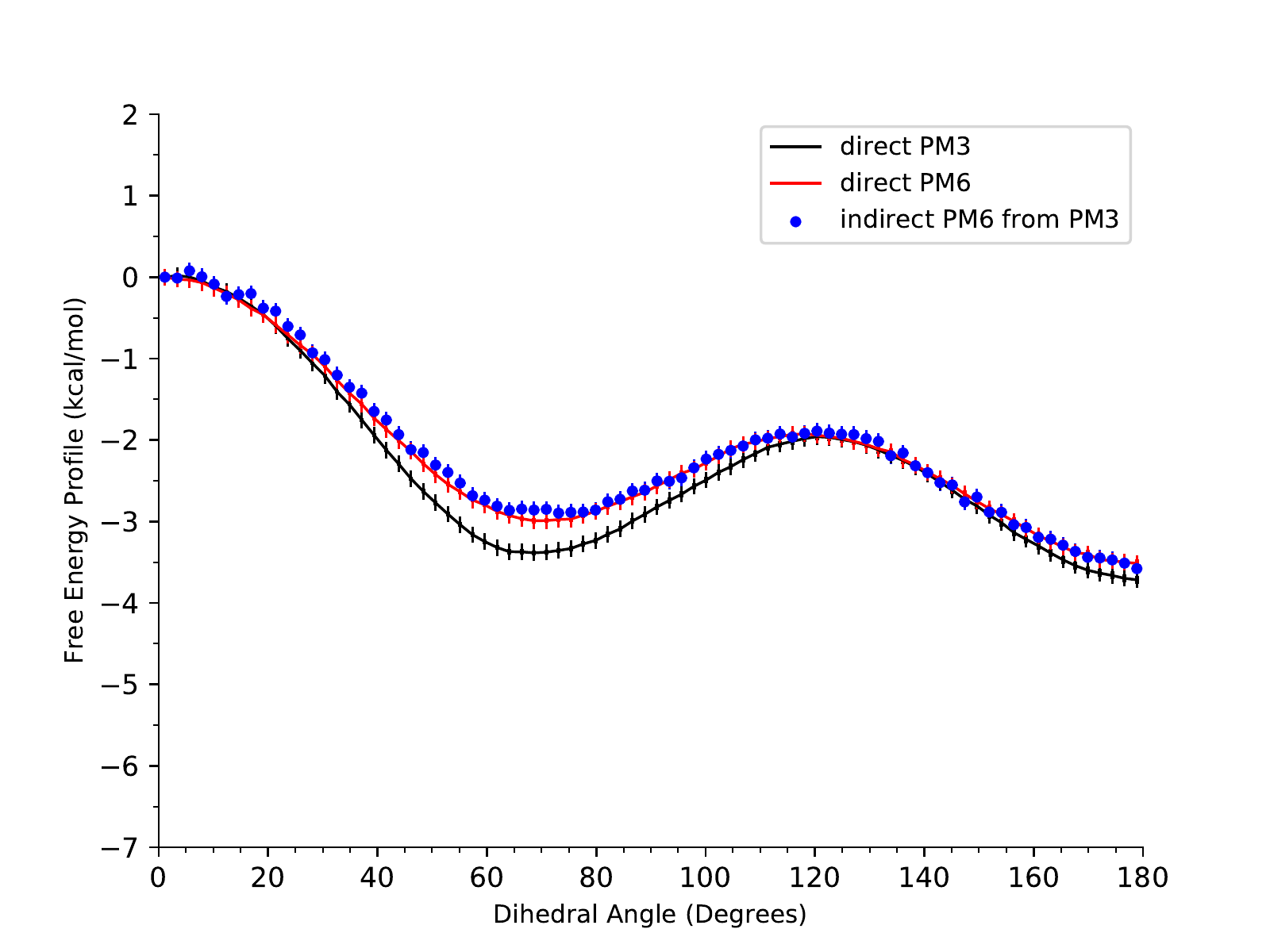}
		\caption{}
		\label{fig:sfig1}
	\end{subfigure}%
	\begin{subfigure}{.5\textwidth}
		\centering
		\includegraphics[width=1.0\linewidth]{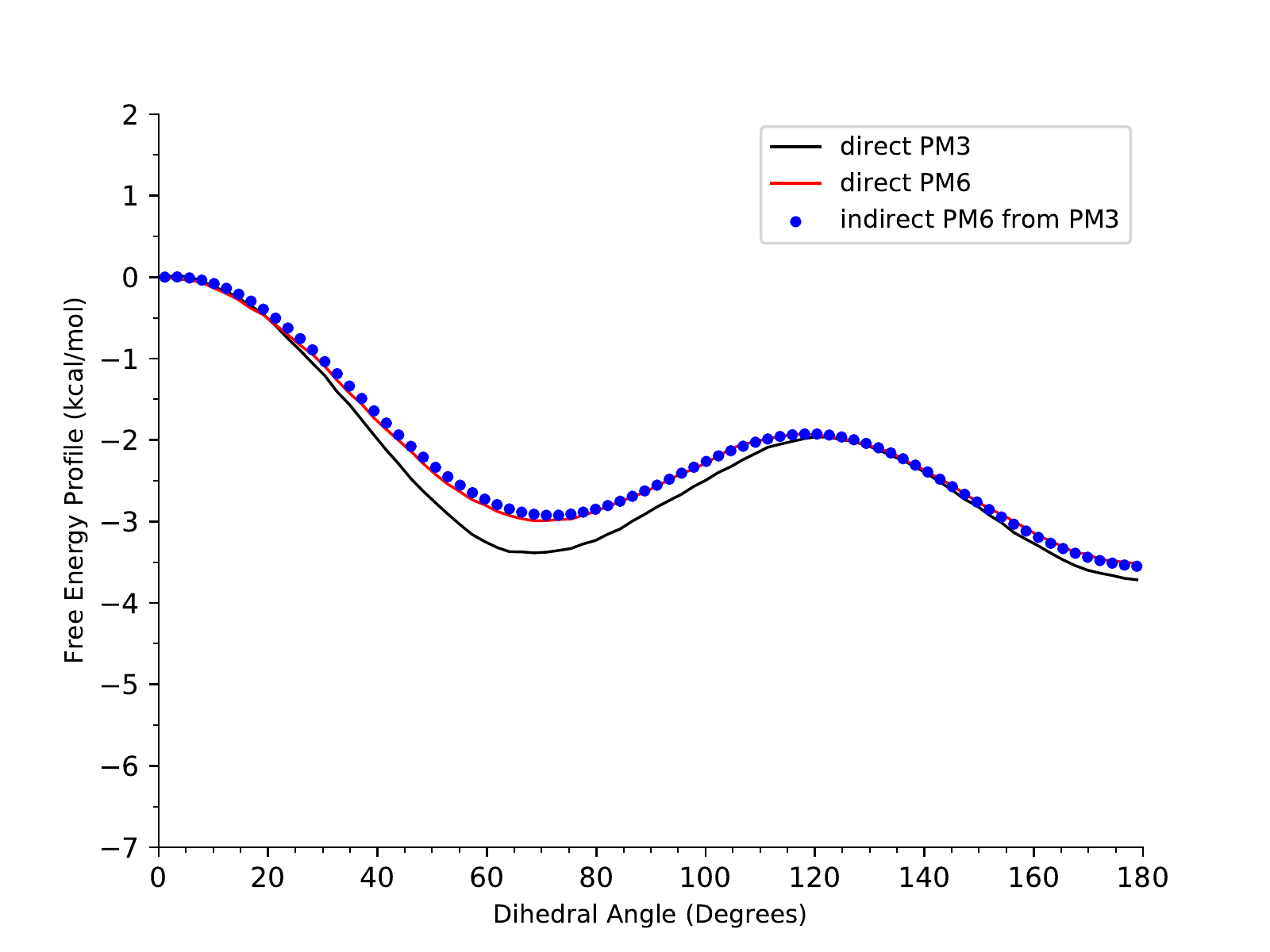}
		\caption{}
		\label{fig:sfig2}
	\end{subfigure}
	\caption{\label{fig:BT:PM3toPM6-PMF-GPR-RE} (a) The direct FE profiles for the main chain dihedral rotation of a butane molecule at the PM3 level (black) and the PM6 level (red), as well as the indirect FE profile at the PM6 level corrected from the PM3 curve (blue dots) before Gaussian process regression smoothing. (b) The indirect profile (blue dots) has been smoothed by Gaussian process regression.}
\end{figure}

Although the minimum free energy structures are located in the same RC value (dihedral angle) under PM3/MM and PM6/MM Hamiltonians, the well depths are different. For the global minimum with $\phi=180^{\circ}$, these two Hamiltonians led to free energy values that differ by 0.2 kcal/mol. The other free energy minimum is located at $\phi=68.5^{\circ}$. \lpfrev{As shown in Table~\ref{tab:FEvalues}}, the well depth from the PM3/MM calculation is -3.39 kcal/mol, which is 0.4 kcal/mol deeper than that from the PM6/MM calculation (-2.99 kcal/mol). 
Although the weighted TP correction was rigorously derived, its practical applicability is still worth a validation. 
In the first step, the PM3-to-PM6 correction was applied to the direct PM3 FE profile, and the indirect PM6 FE profile obtained in this way was compared with the direct PM6 FE profile. As shown in Fig.~\ref{SI-fig:BT-PM3andPM6toB3LYP-RE}, all the reweighting entropy values are larger than 0.76, which indicate that this TP calculation is trustworthy and a smooth indirect FE profile can be obtained. It clearly shows in Fig.~\ref{fig:BT:PM3toPM6-PMF-GPR-RE} that the indirect PM6 FE profile is highly consistent with the direct PM6 FE profile, indicating that the reweighting method introduced in this work is not only theoretically rigorous but also practically reliable at least for the correction from PM3 to PM6.
\begin{table}
	\caption{\label{tab:FEvalues} Free energy well depth of the main chain dihedral rotation of a butane molecule (BT) and free energies barriers for the $\mathrm{S_N2}$ reaction ($\mathrm{S_N2}$), the intramolecular proton transfer reaction of glycine (PT), and the aliphatic Claisen rearrangement reaction (CR). (All the data have a unit of kcal/mol).}
	\newcommand{\rb}[1]{\raisebox{1.5ex}[0pt]{#1}}
	\centering
	\scalebox{1.0}{
		\begin{tabular}{lcccccccccccccc}
			\hline
			&&&\multicolumn{3}{c}{B3LYP} \\
			\cline{4-6}
			\rb{Reaction}       &\rb{direct PM3}&\rb{direct PM6}&indirect from PM3&indirect from PM6&direct\\
			\hline
			BT                  &-3.39$\pm$0.03$^{a}$&-2.99$\pm$0.03$^{a}$&-6.18$\pm$0.05$^{b}$&-5.83$\pm$0.05$^{b}$&-5.86$\pm$0.11$^{b}$ \\
			$\mathrm{S_N2}$ &25.31$\pm$0.07&20.60$\pm$0.07&22.44$\pm$0.08 &22.15$\pm$0.08 &21.70$\pm$0.16 \\
			PT            &24.47$\pm$0.08& 3.25$\pm$0.05&2.79$\pm$0.10&3.67$\pm$0.08&3.04$\pm$0.14 \\
			CR            &35.43$\pm$0.08&30.42$\pm$0.08&27.48$\pm$0.10&28.05$\pm$0.10&26.30$\pm$0.17 \\		
			\hline
		\end{tabular}
	}
	\begin{flushleft}
		\textsuperscript{\emph{a}} Location at first well depth. \\
		\textsuperscript{\emph{b}} Location at $\phi=180^{\circ}$. \\
		%		\textsuperscript{\emph{c}} Activation free energy of reactions. \\
	\end{flushleft}
\end{table}

\begin{figure}
	\begin{subfigure}{.5\textwidth}
		\centering
		\includegraphics[width=1.0\linewidth]{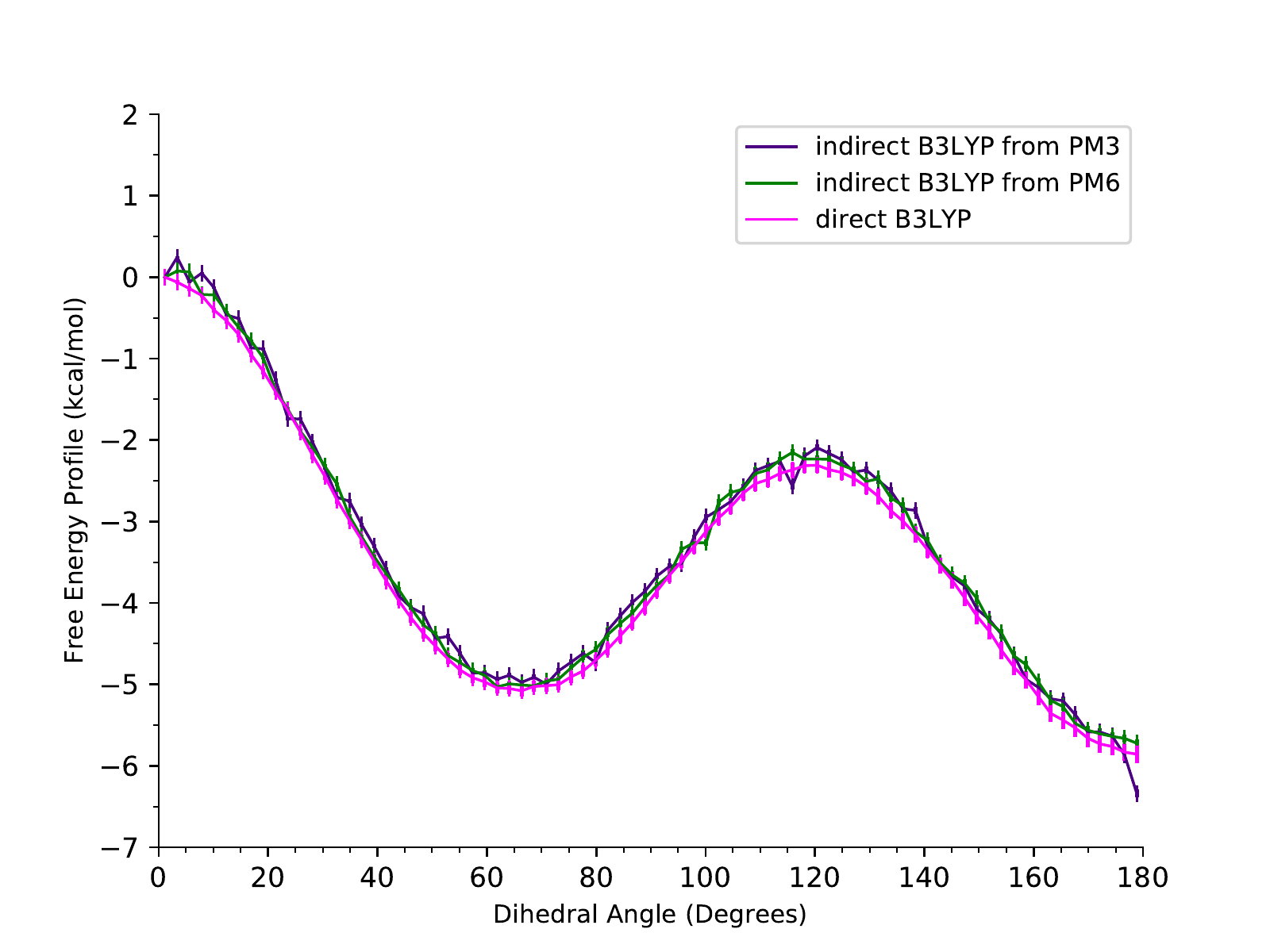}
		\caption{}
		\label{fig:sfig1}
	\end{subfigure}%
	\begin{subfigure}{.5\textwidth}
		\centering
		\includegraphics[width=1.0\linewidth]{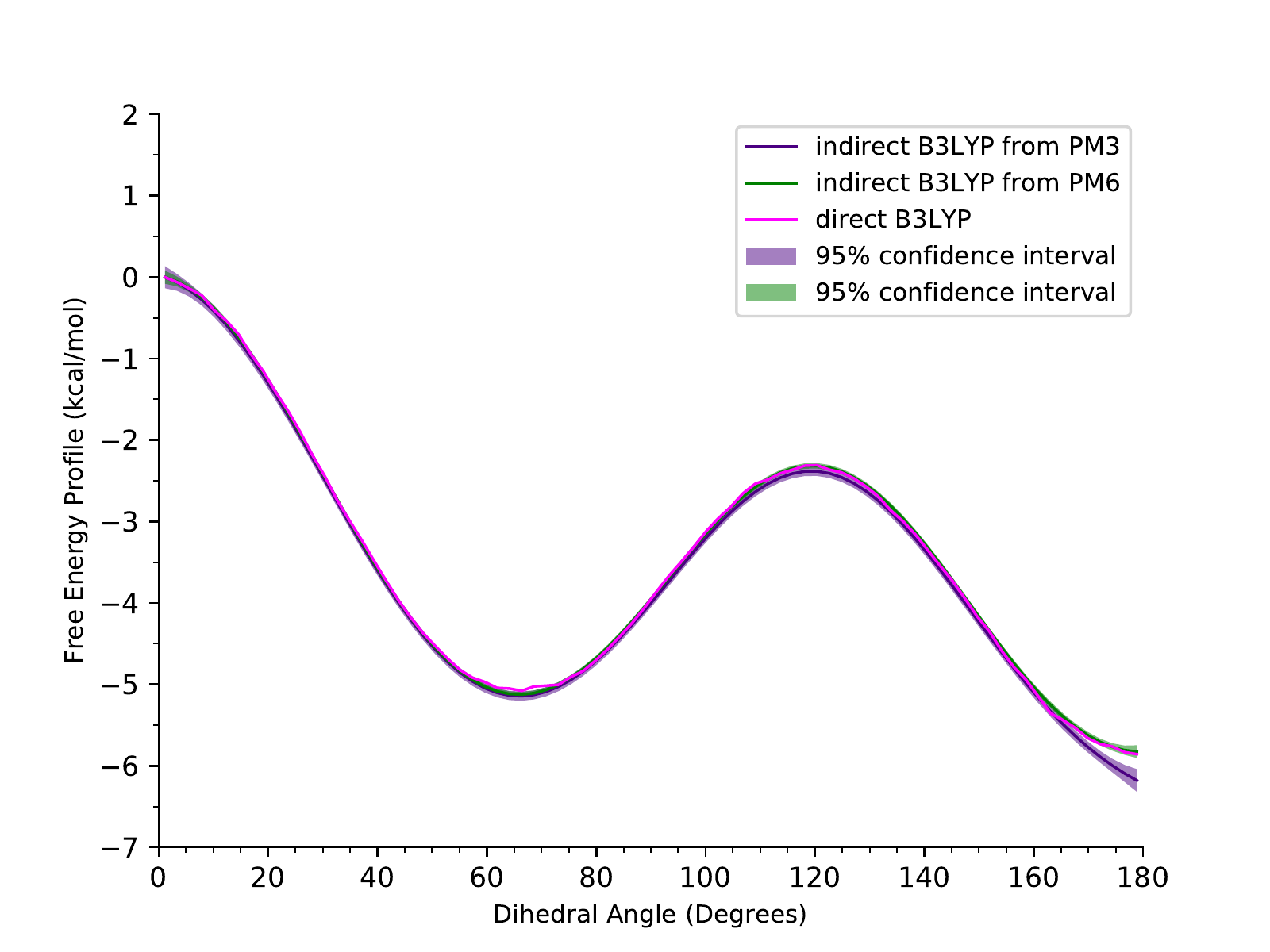}
		\caption{}
		\label{fig:sfig2}
	\end{subfigure}
	\caption{\label{fig:BT:PM3andPM6toB3LYP-PMF-GPR-RE} (a) The indirect FE profiles for the main chain dihedral rotation of a butane molecule at B3LYP level calculated by the weighted TP calculations from the PM3 Hamiltonian (indigo) and the PM6 Hamiltonian (green) before Gaussian process regression smoothing, as well as the direct FE profile at the B3LYP level (magenta). (b) The indirect profiles (indigo and green) have been smoothed by Gaussian process regression. The 95\% confidence intervals are also presented.}
\end{figure}

Next, the PM3-to-B3LYP and PM6-to-B3LYP corrections were applied to generate the indirect FE profiles at the B3LYP level. As shown in Fig.~\ref{SI-fig:BT-PM3andPM6toB3LYP-RE}, the reweighting entropies are smaller than those in the TP calculations from PM3 to PM6, indicating that the overlaps in phase space between B3LYP and the SQM Hamiltonians are less significant than that between PM3 and PM6. Comparing with PM3, the PM6 Hamiltonian is more similar to B3LYP as can be inferred from the reweighting entropy values. This is reasonable, because PM6 was refined over the PM3 Hamiltonian. Nevertheless, most of the weighting entropy values are still larger than 0.6, with only a small number of exceptions. It can be noted that at $\phi=180^{\circ}$, there is a precipitous drop of reweighting entropy in reweighting the PM3 level to the B3LYP level process in Fig.~\ref{SI-fig:BT-PM3andPM6toB3LYP-RE}, which leads to a corresponding precipitous drop of the free energy around this region as shown in Fig.~\ref{fig:BT:PM3andPM6toB3LYP-PMF-GPR-RE}(a). This indicates that the weighted TP corrections for each histogram bin is sensitive to the reweighting entropy quantity, and the reweighting entropy can be used to characterize the reliability of the weighted TP calculations. Then, we fitted these two indirect FE profiles by Gaussian process regression, and the smoothed curves are nearly identical except for a small difference near $\phi=180^{\circ}$, as shown in Fig.~\ref{fig:BT:PM3andPM6toB3LYP-PMF-GPR-RE}(b). They are also very close to the direct FE profile at the same level, and the indirect one from the PM6 level performs slightly better than the indirect one corrected from the PM3 level. \lpfrev{As shown in Table~\ref{tab:FEvalues}}, the two indirect B3LYP FE profiles reweighted from the PM3 and PM6 levels have a free energy of -6.18$\pm$0.05 kcal/mol and -5.83$\pm$0.05 kcal/mol at $\phi=180^{\circ}$, respectively. They well match the direct B3LYP FE profile, which has a free energy of -5.86$\pm$0.11 kcal/mol at $\phi=180^{\circ}$. Meanwhile, all the B3LYP FE profiles, either direct or indirect, have the same positions in RC for the maximum and minimum free energy structures.

\subsection{\label{sec:result:SN2}$\mathrm{\mathbf{S_{N} 2}}$ reaction of \ce{CH3Cl + Cl- -> Cl- + CH3Cl}}  
\begin{figure}
	\begin{subfigure}{.5\textwidth}
		\centering
		\includegraphics[width=1.0\linewidth]{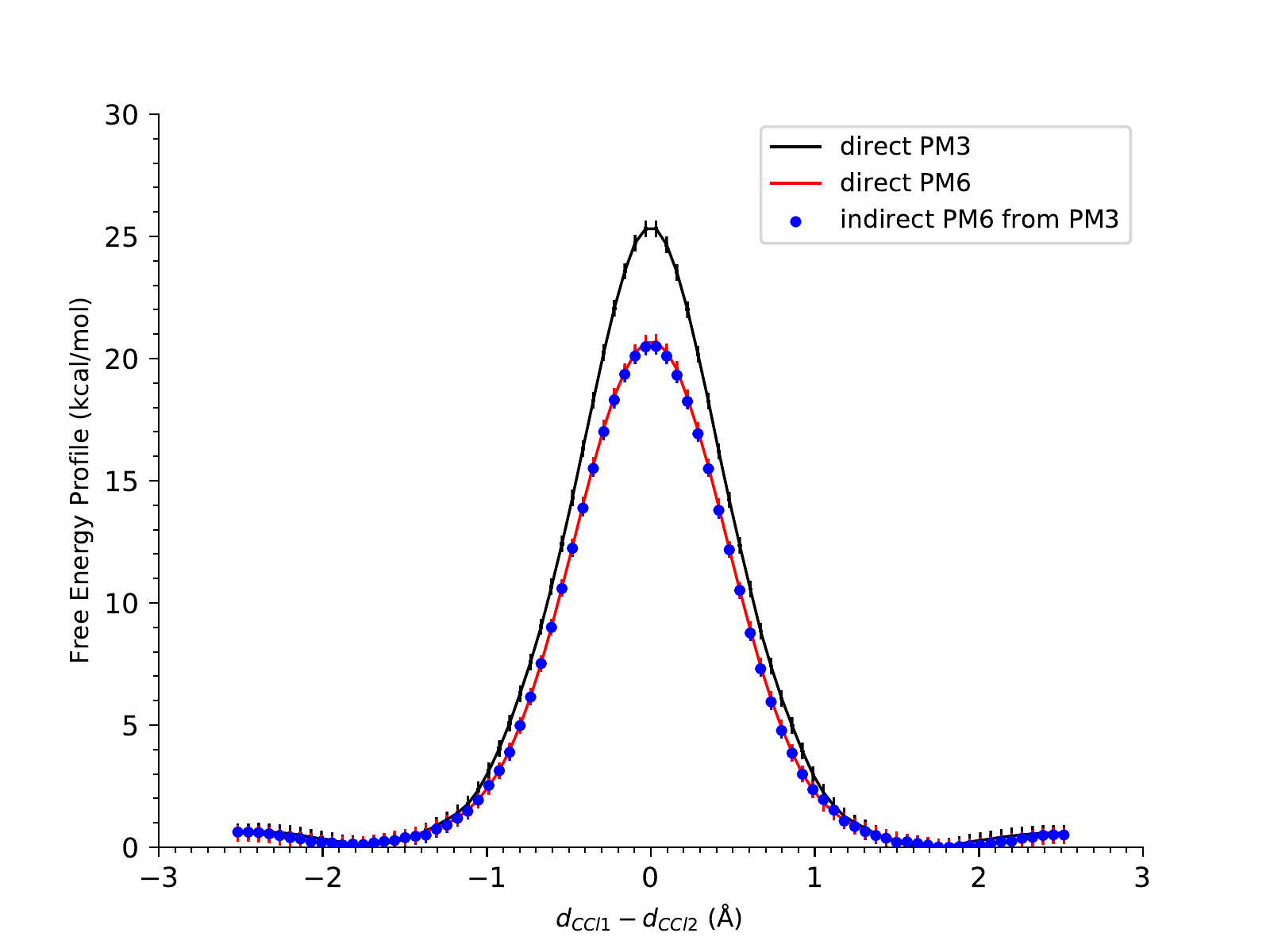}
		\caption{}
		\label{fig:sfig1}
	\end{subfigure}%
	\begin{subfigure}{.5\textwidth}
		\centering
		\includegraphics[width=1.0\linewidth]{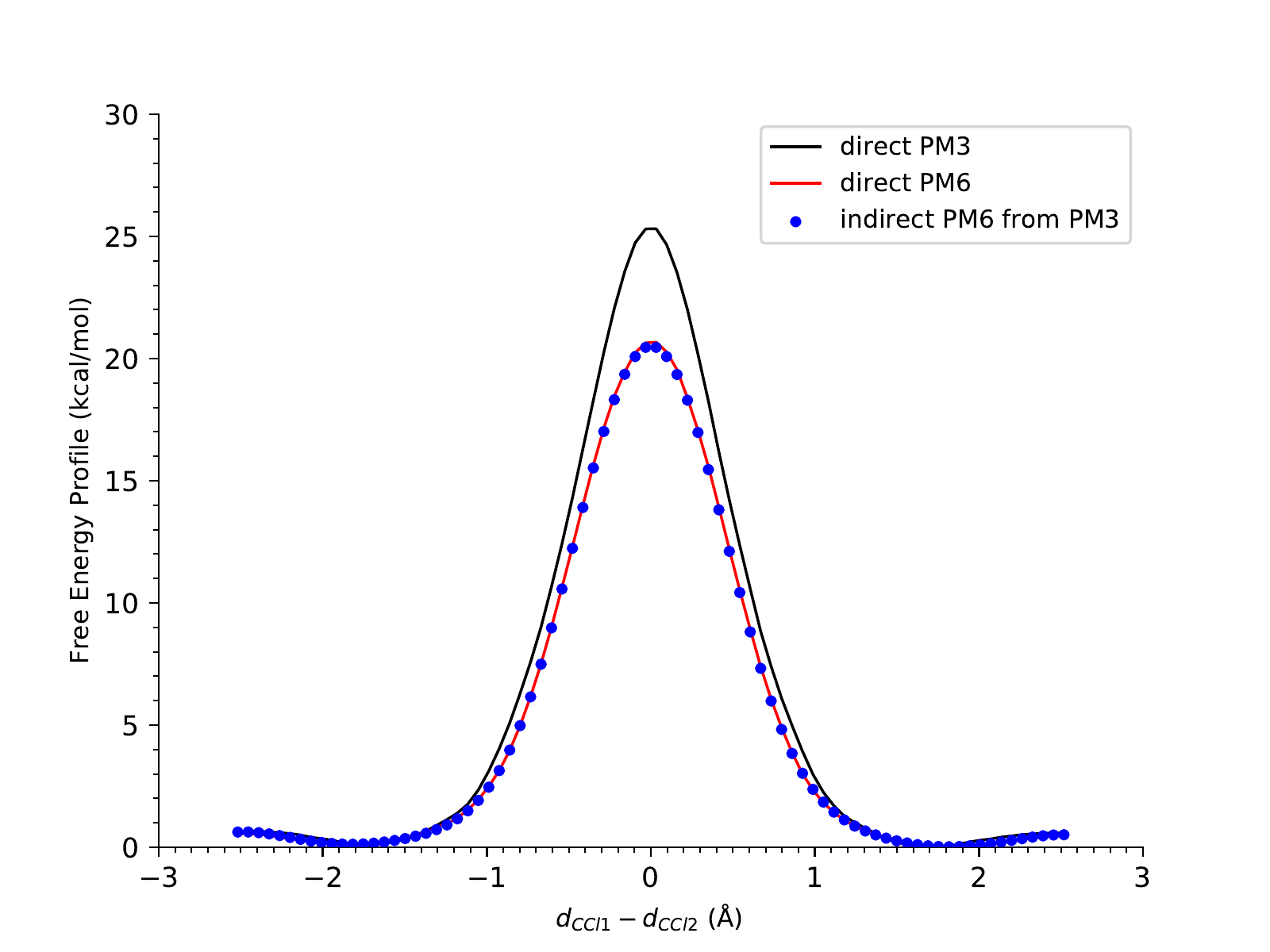}
		\caption{}
		\label{fig:sfig2}
	\end{subfigure}
	\caption{\label{fig:SN2:PM3toPM6-PMF-GPR-RE} (a) The direct FE profiles for the $\mathrm{S_{N} 2}$ reaction of \ce{CH3Cl + Cl- -> Cl- + CH3Cl} at the PM3 level (black) and the PM6 level (red), as well as the indirect FE profile at the PM6 level corrected from the PM3 curve (blue dots) before Gaussian process regression smoothing. (b) The indirect profile (blue dots) has been smoothed by Gaussian process regression.}
\end{figure}

In this work, as shown in Fig.~\ref{fig:SN2:PM3toPM6-PMF-GPR-RE}, the direct FE profile calculated at the PM3 level shows some difference from that at the PM6 level. \lpfrev{As shown in Table~\ref{tab:FEvalues}}, the free energy barrier estimated from the PM3 simulations is about 4.7 kcal/mol higher than that obtained from the PM6 simulations. The element-specific Gaussian core-core corrections in PM3 is replaced with a pairwise core-core correction term in PM6.\cite{StewartJMM2007,ChristensenCR2016} Besides, $d$-orbitals are added to the atomic basis for certain elements including chlorine in the PM6 Hamiltonian, which shows significant improvement over the PM3 method.\cite{ThielJPC1996} Therefore, it is expected that PM6 Hamiltonian outperforms PM3 for the agreement of the FE profile with the one at the B3LYP level. After the PM3-to-PM6 correction, the indirect PM6 FE profile is highly consistent with the direct PM6 FE profile, as shown in Fig.~\ref{fig:SN2:PM3toPM6-PMF-GPR-RE}. The reweighting entropies are all larger than 0.70 (see Fig.~\ref{SI-fig:SN2-PM3andPM6toB3LYP-RE}), which also indicates that this TP calculation can generate a reliable indirect FE curve.

\begin{figure}
	\begin{subfigure}{.5\textwidth}
		\centering
		\includegraphics[width=1.0\linewidth]{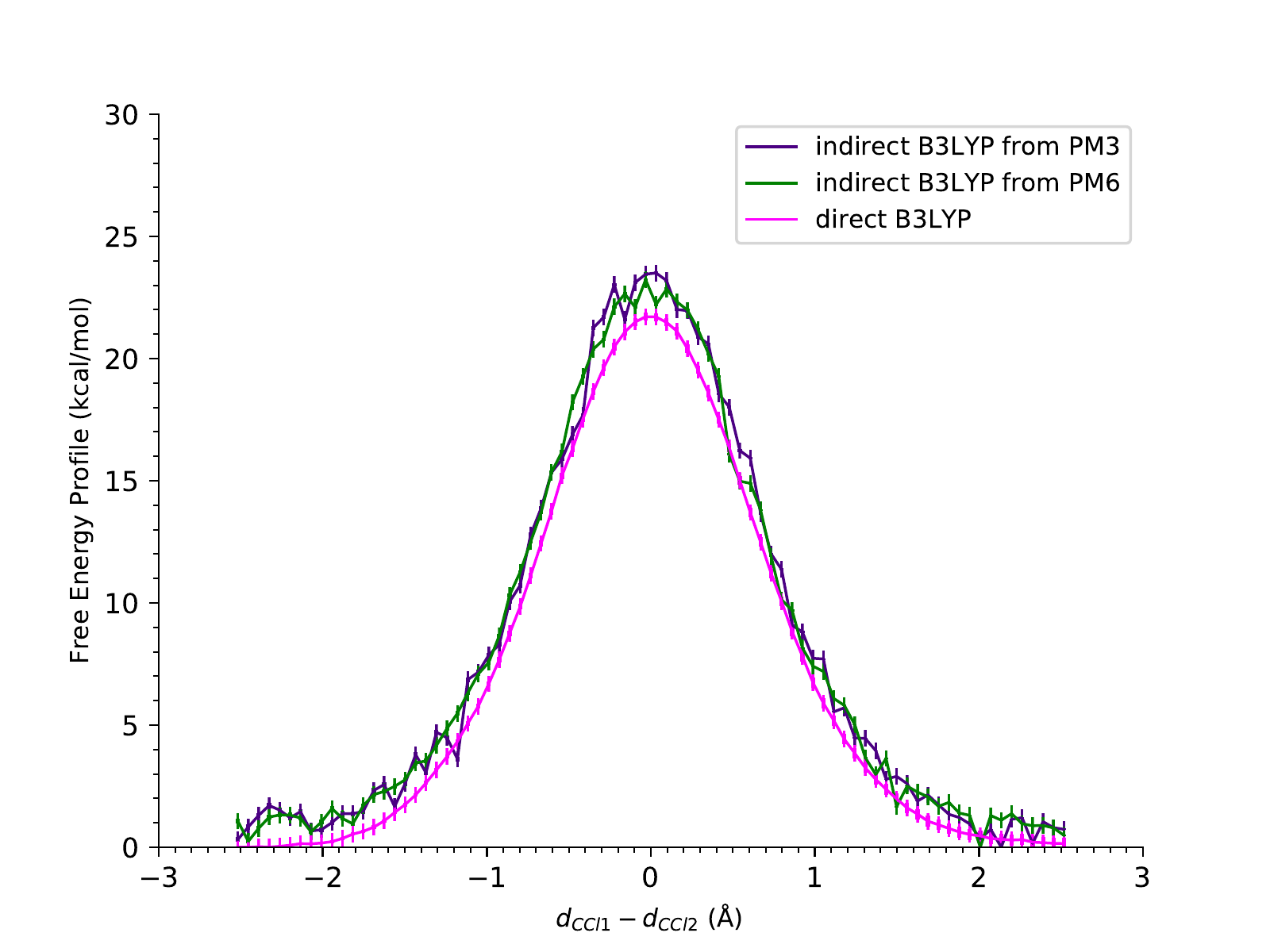}
		\caption{}
		\label{fig:sfig1}
	\end{subfigure}%
	\begin{subfigure}{.5\textwidth}
		\centering
		\includegraphics[width=1.0\linewidth]{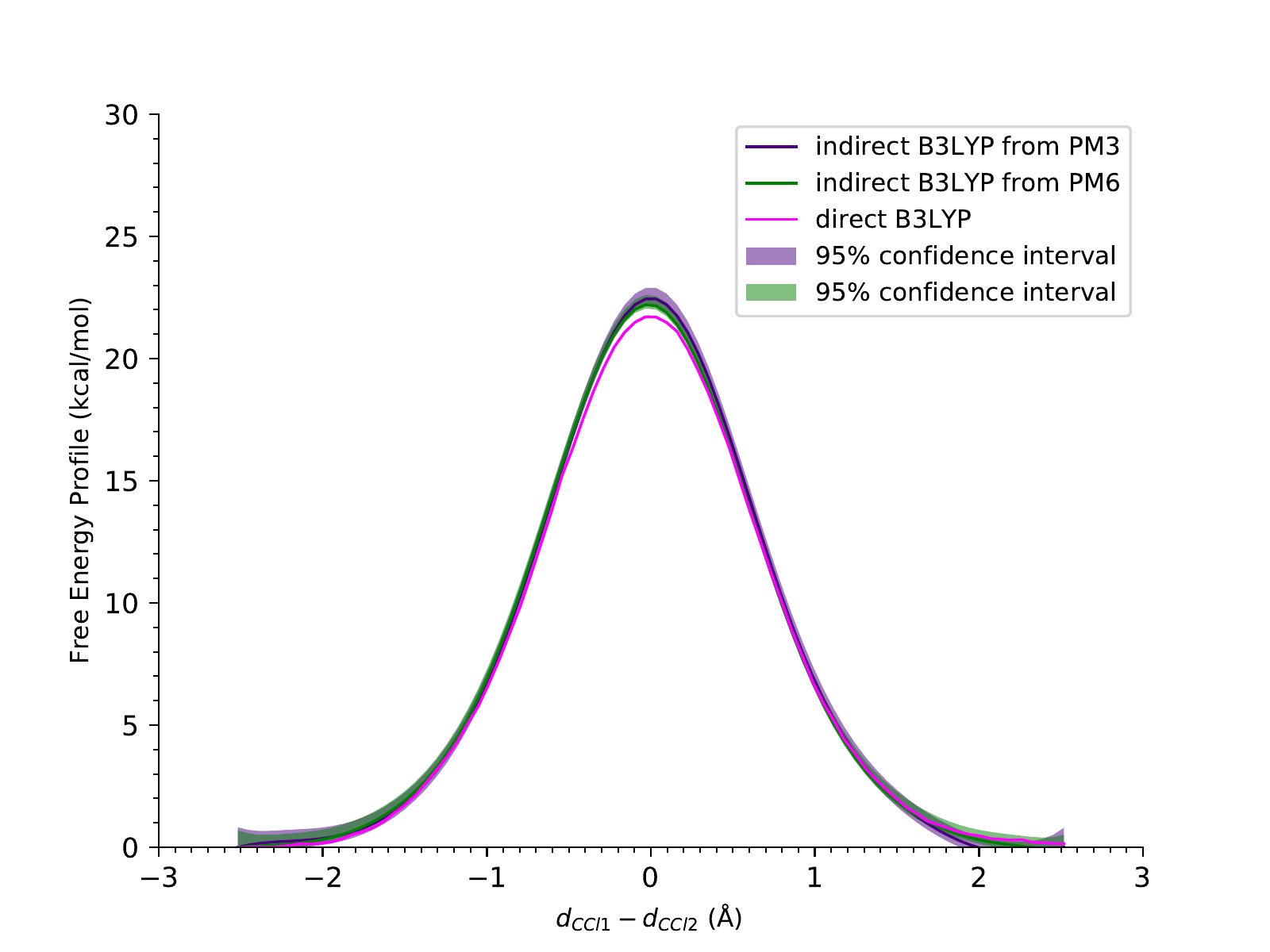}
		\caption{}
		\label{fig:sfig2}
	\end{subfigure}
	\caption{\label{fig:SN2:PM3andPM6toB3LYP-PMF-GPR-RE} (a) The indirect FE profiles for the $\mathrm{S_{N} 2}$ reaction of \ce{CH3Cl + Cl- -> Cl- + CH3Cl} at the B3LYP level calculated by the weighted TP calculations from the PM3 Hamiltonian (indigo) and the PM6 Hamiltonian (green) before Gaussian process regression smoothing, as well as the direct FE profile at the B3LYP level (magenta). (b) The indirect profiles (indigo and green) have been smoothed by Gaussian process regression. The 95\% confidence intervals are also presented.}
\end{figure}

However, when running the TP calculations from the SQM to B3LYP level, the reweighting entropies became much smaller than those in the PM3-to-PM6 TP corrections (see Fig.~\ref{SI-fig:SN2-PM3andPM6toB3LYP-RE}). This is caused by the slow convergence at the tail region of the biasing potential distribution.\cite{JiaJCTC2016,HeimdalPCCP2012,NBBJCTC,Cave-AylandJPCB2015,KlimovichJCAMD2015,LuFECbook,RydeJCTC2017} It is not surprising that the fitted FE profile data are noisy, which are shown in Fig.~\ref{fig:SN2:PM3andPM6toB3LYP-PMF-GPR-RE}(a). This statistical fluctuation can be well smoothed by the Gaussian process regression, and the final curves are nearly identical, as shown in Fig.~\ref{fig:SN2:PM3andPM6toB3LYP-PMF-GPR-RE}(b). In addition, these two indirect B3LYP FE profiles are highly consistent with the direct B3LYP FE profile, except for some marginal difference at the transition state and product state regions. Especially, when corrected from the PM6 level, the transition barrier differs by only 0.45 kcal/mol from the direct FE profile at the B3LYP level, \lpfrev{as shown in Table~\ref{tab:FEvalues}}.

Please note that the simulated reaction barrier is around 22 kcal/mol, which is lower by 4-5 kcal/mol than the experimentally measured one (26.5 kcal/mol) determined from rate constant. This deviation is likely to be caused by the limited accuracy of B3LYP functional and the approximate QM/MM van der Waals interaction, which are not fully compatible with each other. The agreement is expected to be further improved when a more accurate description is adopted for both the intra-molecular and inter-molecular interactions. For instance, Kuechler et al studied this reaction using a highly accurate specific reaction parameter (SRP) semi-empirical method\cite{KuechlerJCP2014} and a charge-dependent exchange and dispersion (QXD) model\cite{KuechlerJCP2015}, and found excellent agreement between the simulated and experimental barriers for this reaction. However, the accuracy of the high-level Hamiltonian (B3LYP) is not a focus in the current work.

\subsection{\label{sec:result:PT}Glycine intramolecular proton transfer reaction}
\begin{figure}
	\begin{subfigure}{.5\textwidth}
		\centering
		\includegraphics[width=1.0\linewidth]{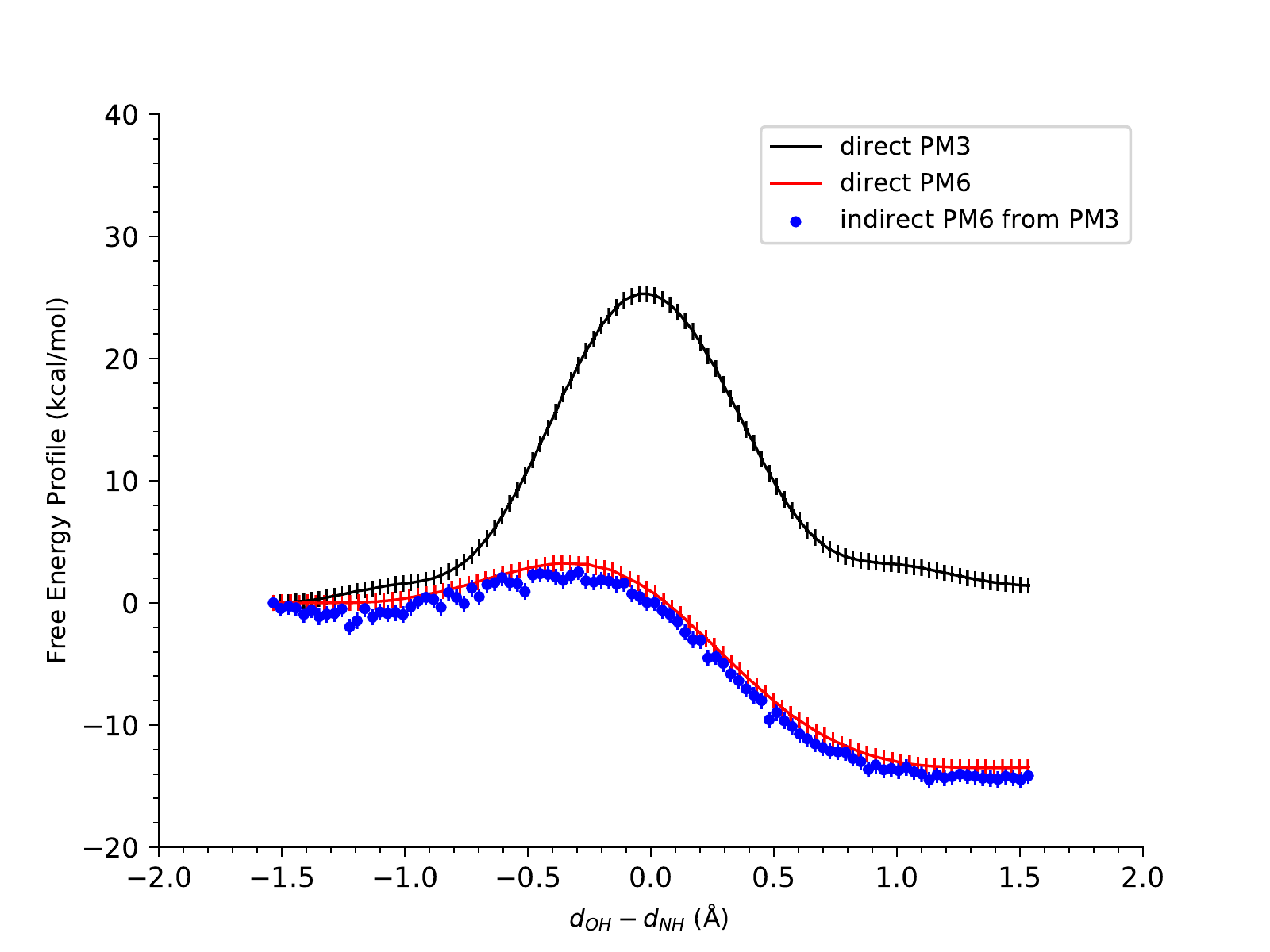}
		\caption{}
		\label{fig:sfig1}
	\end{subfigure}%
	\begin{subfigure}{.5\textwidth}
		\centering
		\includegraphics[width=1.0\linewidth]{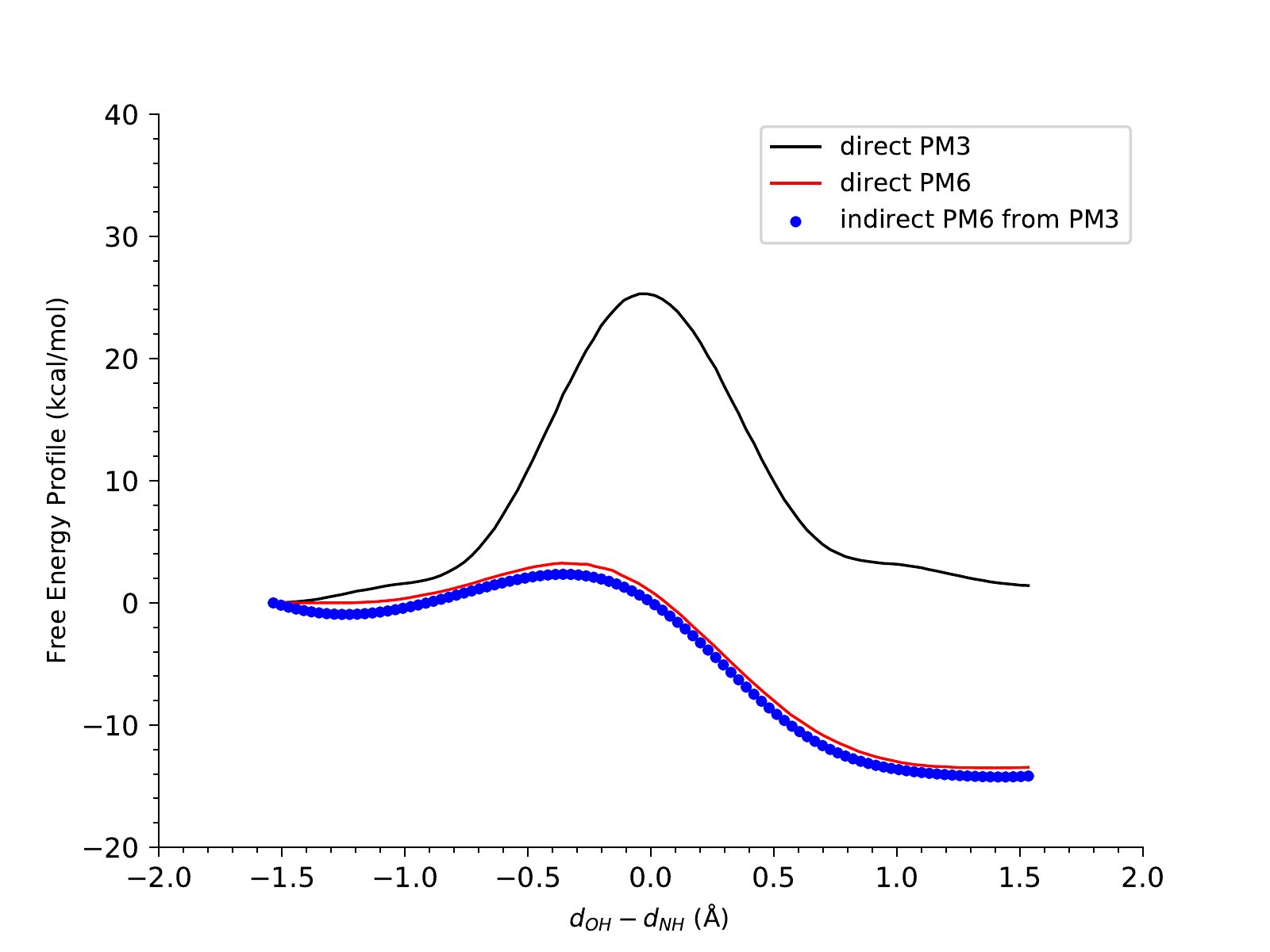}
		\caption{}
		\label{fig:sfig2}
	\end{subfigure}
	\caption{\label{fig:PT:PM3toPM6-PMF-GPR-RE} (a) The direct FE profiles for the glycine intramolecular proton transfer reaction at the PM3 level (black) and the PM6 level (red), as well as the indirect FE profile at the PM6 level corrected from the PM3 curve (blue dots) before Gaussian process regression smoothing. (b) The indirect profile (blue dots) has been smoothed by Gaussian process regression.}
\end{figure}

As shown in Fig.~\ref{fig:PT:PM3toPM6-PMF-GPR-RE}, the FE profile calculated at the PM3 level is very different from that at the PM6 level. \lpfrev{As shown in Table~\ref{tab:FEvalues},} the reaction free energy and activation barrier estimated from the PM3/MM MD simulations are 1.51 $\pm$ 0.11 and 24.47 $\pm$ 0.08 kcal/mol, respectively, while those obtained from the PM6/MM simulations are -13.45 $\pm$ 0.08 and 3.25 $\pm$ 0.05 kcal/mol. The transition state at PM3 level is located at -0.02 {\AA}, while that at PM6 level is located at -0.36 {\AA}. Clearly, the PM3/MM simulations yielded an incorrect result, because the zwitterion form of glycine molecule is stabilized by the water molecules and is the dominant state in aqueous solution. The glycine intramolecular proton transfer reaction process involves one hydrogen atom transferred from the oxygen atom to the nitrogen atom. Because PM6 uses different core-core repulsion potentials for \ce{N-H}, \ce{O-H}, \ce{C-C}, \ce{Si-O} pairs to correct for the specific defect in the parametrization, it is not surprising that it yielded a FE profile that is more reasonable than the PM3 one. 
As is shown in Fig.~\ref{SI-fig:PT-PM3andPM6toB3LYP-RE}, some of the reweighting entropies are rather small in magnitude in comparison with those obtained for the two systems above, especially for the bins with RC from -1.5 {\AA} to -0.5 {\AA}. After the TP correction, the indirect PM6 FE profile shows some deviations from the direct one in this region. It is interesting to see that, after the Gaussian process regression, the indirect PM6 FE profile can be well superimposed with the direct PM6 FE profile. This result demonstrated that this TP correction can even work well despite a large difference in two semiempirical Hamiltonians. 

\begin{figure}
	\begin{subfigure}{.5\textwidth}
		\centering
		\includegraphics[width=1.0\linewidth]{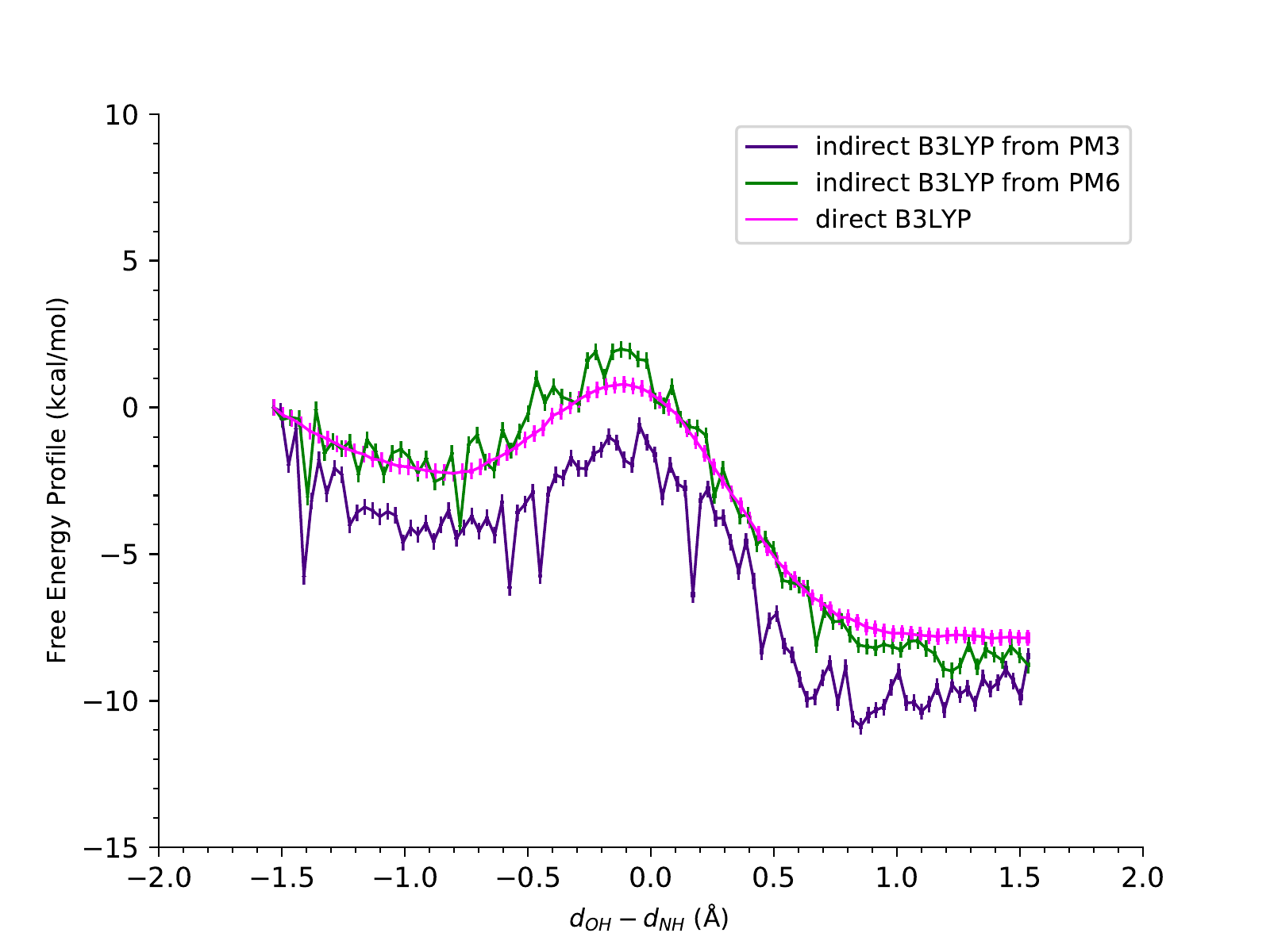}
		\caption{}
		\label{fig:sfig1}
	\end{subfigure}%
	\begin{subfigure}{.5\textwidth}
		\centering
		\includegraphics[width=1.0\linewidth]{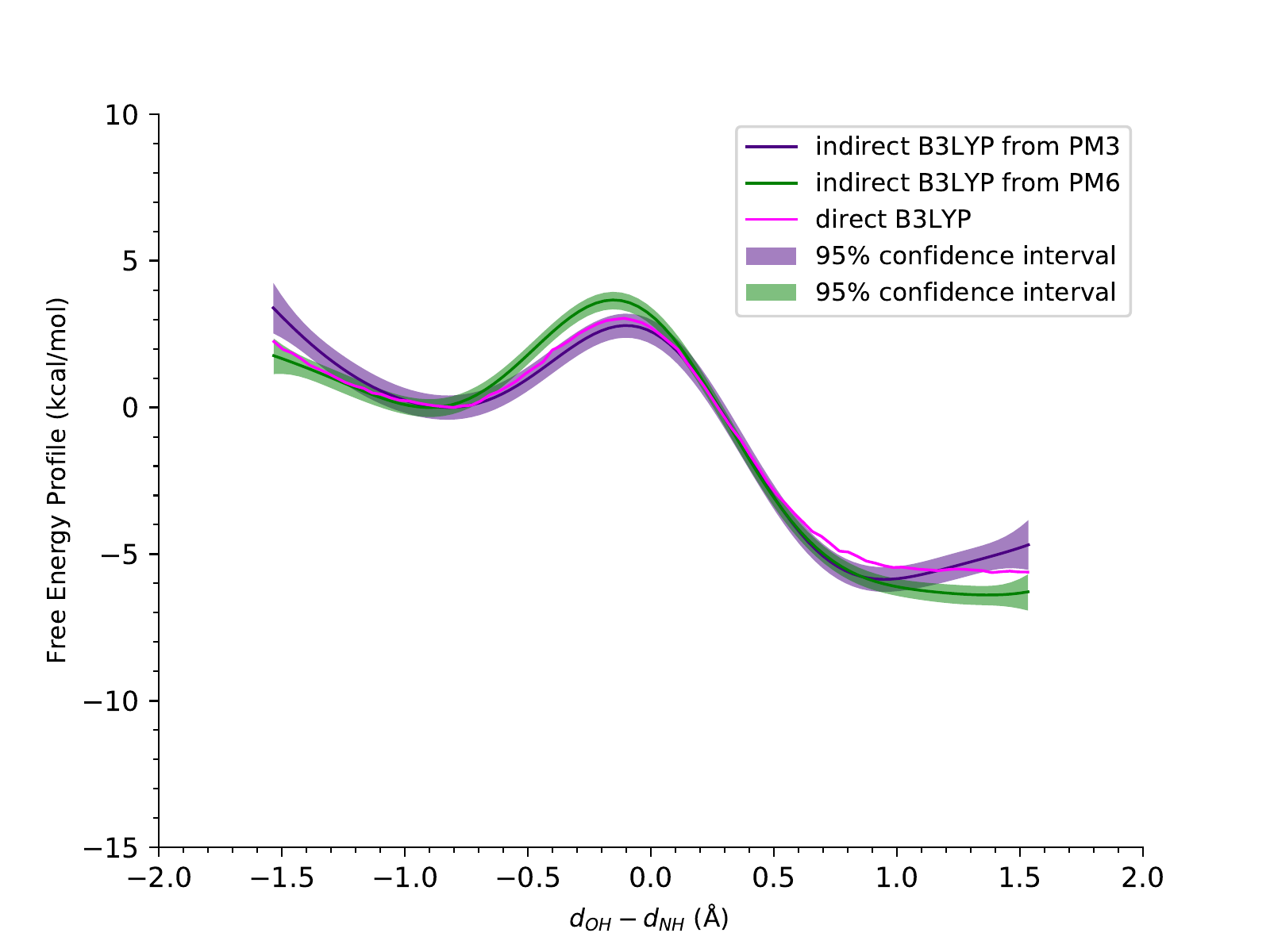}
		\caption{}
		\label{fig:sfig2}
	\end{subfigure}
	\caption{\label{fig:PT:PM3andPM6toB3LYP-PMF-GPR-RE} (a) The indirect FE profiles for the glycine intramolecular proton transfer reaction at the B3LYP level calculated by the weighted TP calculations from the PM3 Hamiltonian (indigo) and the PM6 Hamiltonian (green) before Gaussian process regression smoothing, as well as the direct FE profile at the B3LYP level (magenta). (b) The indirect profiles (indigo and green) have been smoothed by Gaussian process regression. The 95\% confidence intervals are also presented.}
\end{figure}

The reweighting entropies are even smaller in the TP calculations from the PM3 and PM6 levels to the B3LYP level as shown in Fig.~\ref{SI-fig:PT-PM3andPM6toB3LYP-RE}, which leads to even more noisy FE profiles as shown in Fig.~\ref{fig:PT:PM3andPM6toB3LYP-PMF-GPR-RE}(a). Since PM3 Hamiltonian is less accurate than PM6, it is quite natural that the indirect FE profile from PM3 has much more notable fluctuations than the one from PM6. \lpfrev{Among all the reactions studied in this work, glycine intramolecular proton transfer reaction is the most challenging one with many reweighting entropy values below 0.3. This will definitely lead to large uncertainty in the indirect free energy profiles at the B3LYP level. For the TP calculations from PM6 to B3LYP, most of the reweighting entropies are larger than 0.3 when the reaction coordinate $\eta$ is larger than 0 {\AA} (see Fig.~\ref{SI-fig:PT-PM3andPM6toB3LYP-RE}). Correspondingly, the free energy profile in this region is also smooth with only small fluctuations as shown in Fig.~\ref{fig:PT:PM3andPM6toB3LYP-PMF-GPR-RE}(a). While in the region with $-1\,\mathrm{{\AA}}<\eta< 0\, \mathrm{{\AA}}$, many points have a reweighting entropy smaller than 0.3. The fluctuations of the free energy in this region are more significant. In Gaussian process regression, we set the variance of data noise to be inversely proportional to the exponential of the reweighting entropy. Data with large reweighting entropy have large weight in determining the profile. While those with small reweighting entropy are less important. Provided that there are some data points with large reweighting entropies (larger than 0.3 for this case) in a small interval, a confident free energy profile can still be generated locally.}
After Gaussian process regression, the curves agree much better with each other, as well as with the direct FE profile at the B3LYP level as shown in Fig.~\ref{fig:PT:PM3andPM6toB3LYP-PMF-GPR-RE}(b). The transition states after TP correction shift to the left by 0.09 {\AA} (at -0.11 {\AA}) for PM3 and to the right by 0.2 {\AA} (at -0.16 {\AA}) for PM6 as compared to the original SQM results. The transition state in the direct B3LYP FE profile locates at -0.11 {\AA}. \lpfrev{As shown in Table~\ref{tab:FEvalues},} the free energy barrier heights in the indirect FE profiles are 2.79 kcal/mol and 3.67 kcal/mol, respectively, which are close to the direct B3LYP FE value (3.04 kcal/mol) with a somewhat acceptable difference. 

\subsection{\label{sec:result:CR}Aliphatic Claisen rearrangement reaction of allyl vinyl ether to 4-pentenal}
\begin{figure}
	\begin{subfigure}{.5\textwidth}
		\centering
		\includegraphics[width=1.0\linewidth]{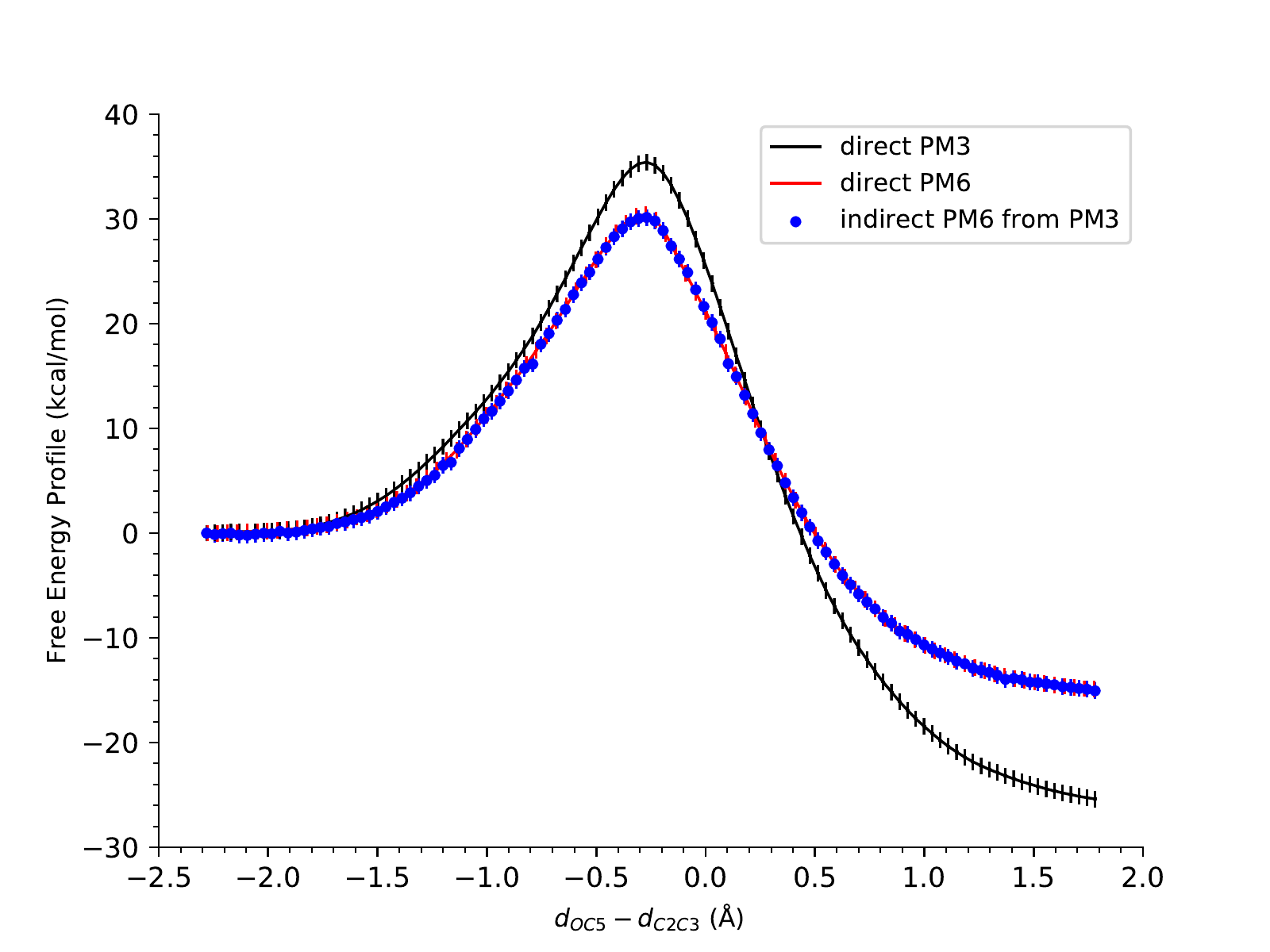}
		\caption{}
		\label{fig:sfig1}
	\end{subfigure}%
	\begin{subfigure}{.5\textwidth}
		\centering
		\includegraphics[width=1.0\linewidth]{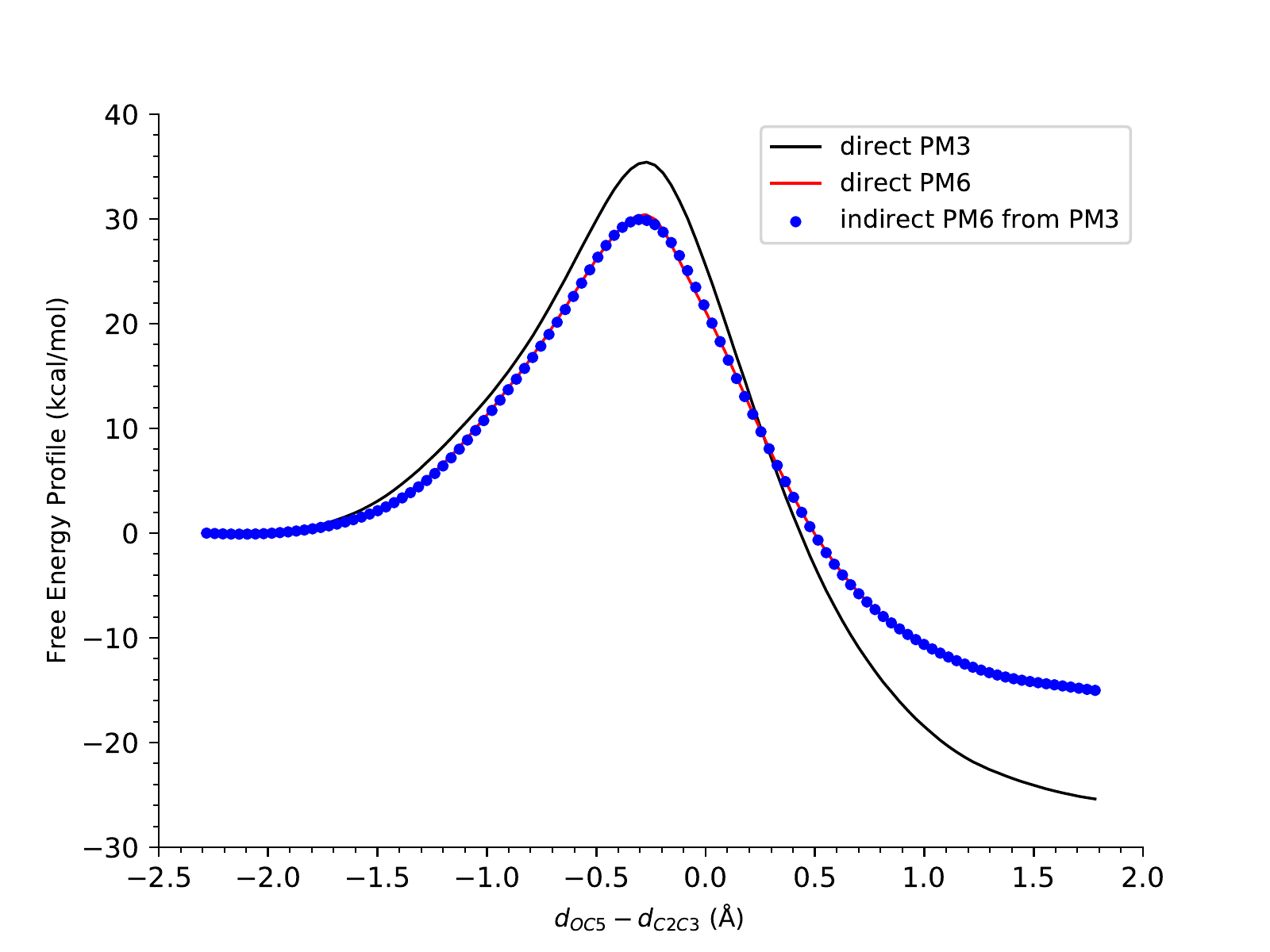}
		\caption{}
		\label{fig:sfig2}
	\end{subfigure}
	\caption{\label{fig:CR:PM3toPM6-PMF-GPR-RE} (a) The direct FE profiles for the aliphatic Claisen rearrangement reaction of allyl vinyl ether to 4-pentenal at the PM3 level (black) and the PM6 level (red), as well as the indirect FE profile at the PM6 level corrected from the PM3 curve (blue dots) before Gaussian process regression smoothing. (b) The indirect profile (blue dots) has been smoothed by Gaussian process regression.}
\end{figure}

Just like the other three reactions, the PM3 FE profile shows some difference from the PM6 one. The reaction barrier is about 5.0 kcal/mol higher (30.42$\pm$0.08 kcal/mol for PM6 v.s. 35.43$\pm$0.08 kcal/mol for PM3), and the reaction free energy is 10.5 kcal/mol more negative (-14.88$\pm$0.11 kcal/mol for PM6 vs -25.39$\pm$0.12 kcal/mol for PM3), \lpfrev{as shown in Table~\ref{tab:FEvalues}}. 
However, the TP correction can still eliminate this difference, and the direct and indirect PM6 FE profiles agree with each other quite well with only marginal difference in some small region as shown in Fig.~\ref{fig:CR:PM3toPM6-PMF-GPR-RE}(a), which was caused mainly by small magnitude of overlaps in the phase space between the high-level and low-level Hamiltonians and can be readily flagged by the reweighting entropy in these regions as shown in Fig.~\ref{SI-fig:CR-PM3andPM6toB3LYP-RE}. Most of the bins have a reweighting entropy larger than 0.6, which guaranteed reliable free energy corrections for those bins. After smoothing using the Gaussian process regression, the direct and indirect FE profiles at PM6 level perfectly agree with each other, as shown in Fig.~\ref{fig:CR:PM3toPM6-PMF-GPR-RE}(b).

\begin{figure}
	\begin{subfigure}{.5\textwidth}
		\centering
		\includegraphics[width=1.0\linewidth]{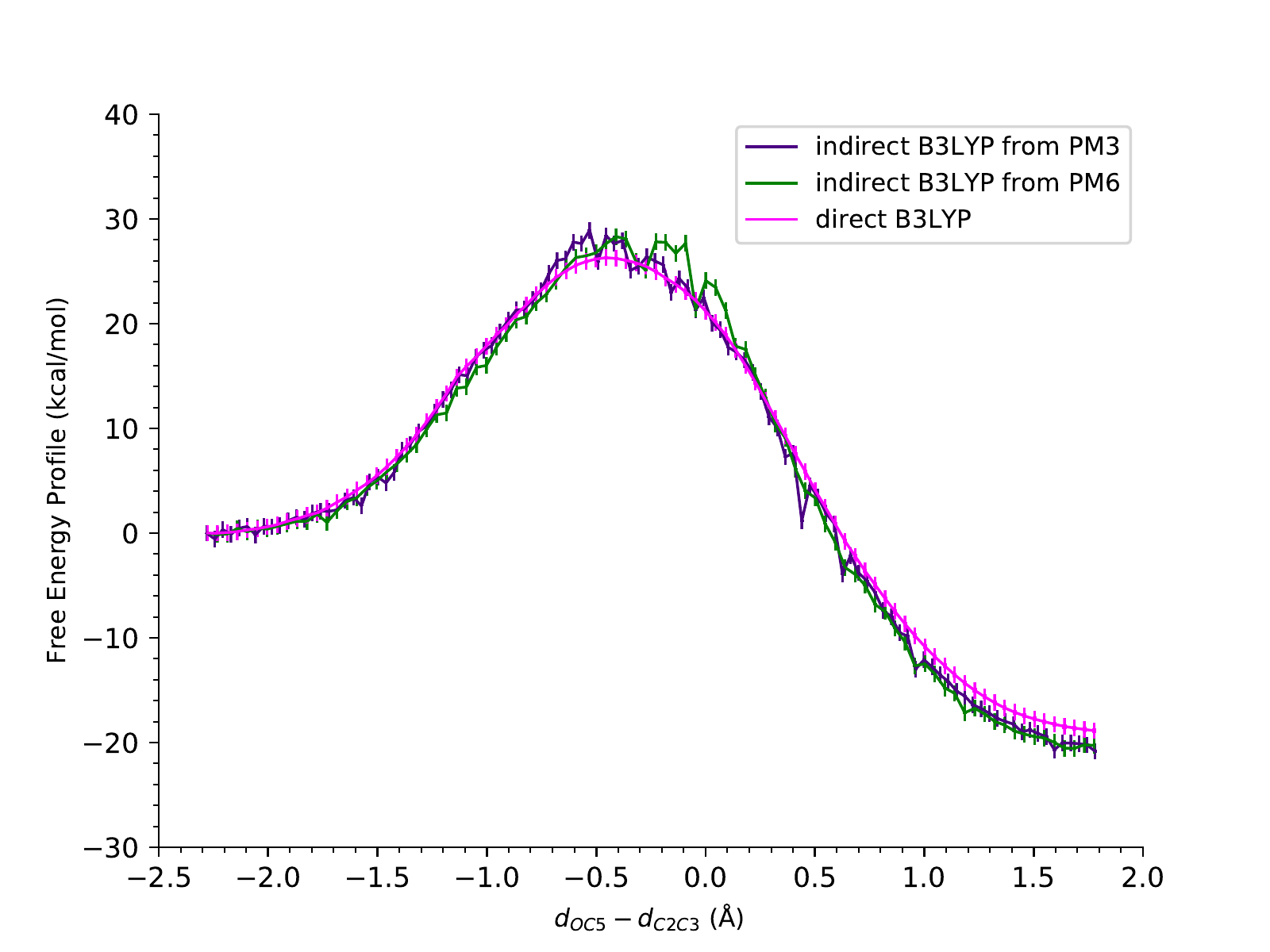}
		\caption{}
		\label{fig:sfig1}
	\end{subfigure}%
	\begin{subfigure}{.5\textwidth}
		\centering
		\includegraphics[width=1.0\linewidth]{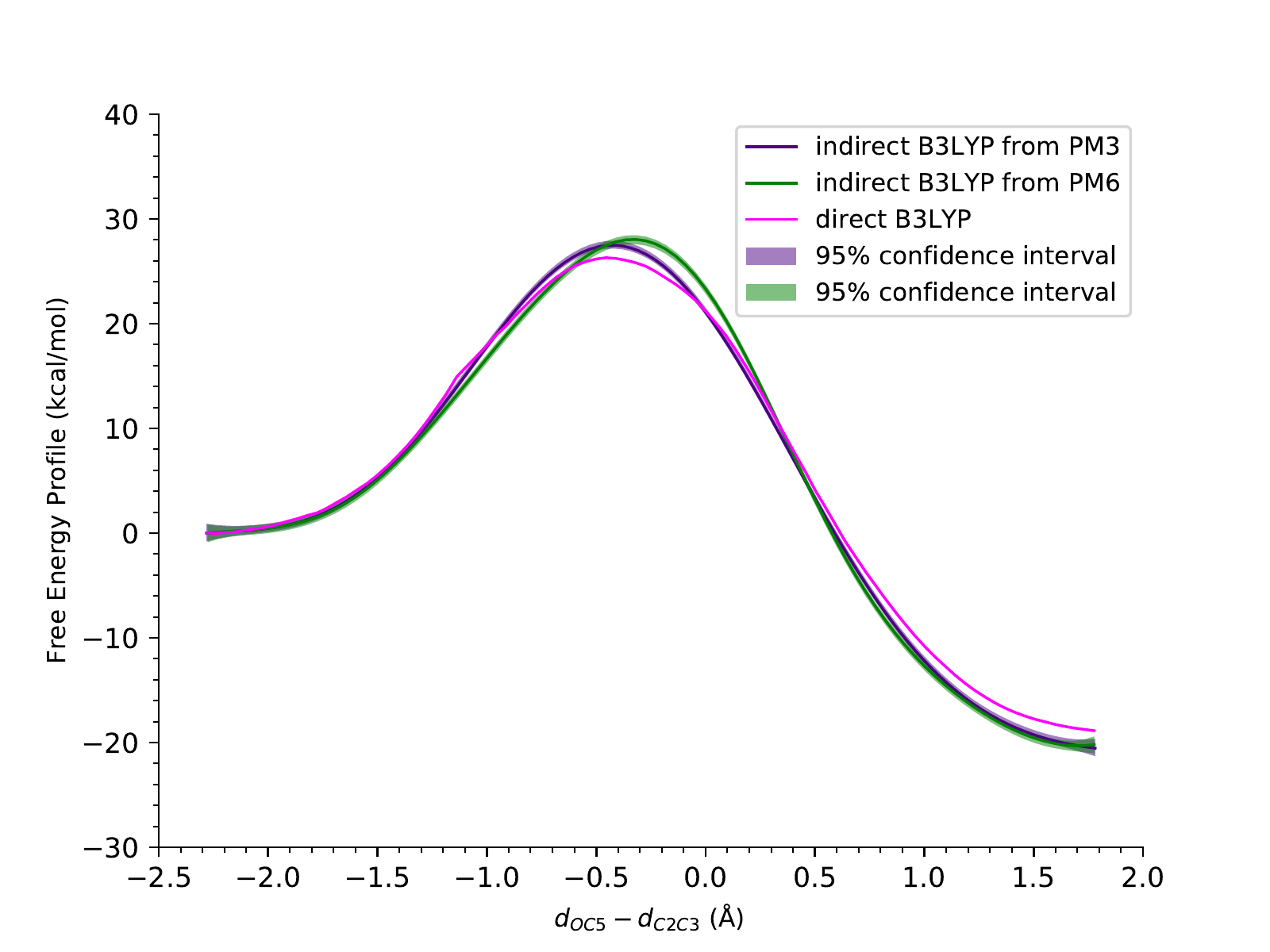}
		\caption{}
		\label{fig:sfig2}
	\end{subfigure}
	\caption{\label{fig:CR:PM3andPM6toB3LYP-PMF-GPR-RE} (a) The indirect FE profiles for the aliphatic Claisen rearrangement reaction of allyl vinyl ether to 4-pentenal at the B3LYP level calculated by the weighted TP calculations from the PM3 Hamiltonian (indigo) and the PM6 Hamiltonian (green) before Gaussian process regression smoothing, as well as the direct FE profile at the B3LYP level (magenta). (b) The indirect profiles (indigo and green) have been smoothed by Gaussian process regression. The 95\% confidence intervals are also presented.}
\end{figure}

In the TP calculations from the PM3 and PM6 levels to the B3LYP level, the reweighting entropies are even smaller, resulting in two rough indirect FE profiles at the B3LYP level shown in Fig.~\ref{fig:CR:PM3andPM6toB3LYP-PMF-GPR-RE}. For the indirect FE profile from PM6, the noise is rather intense in the region with the RC around 0.0 {\AA}, which corresponds well with the small reweighting entropy in that region as shown in Fig.~\ref{SI-fig:CR-PM3andPM6toB3LYP-RE}. The two indirect B3LYP FE profiles after being smoothed by the Gaussian process regression method are highly consistent with each other again, which are shown in Fig.~\ref{fig:CR:PM3andPM6toB3LYP-PMF-GPR-RE}(b).
The transition states after the TP corrections from the PM3 and PM6 Hamiltonians are located at -0.42 {\AA} and -0.32 {\AA}, respectively, which are shifted to the left by 0.15 {\AA} and 0.05 {\AA} as compared to the original SQM results. The transition state in the direct B3LYP/MM FE profile locates at -0.46 {\AA}. \lpfrev{As shown in Table~\ref{tab:FEvalues},} the free energy barrier height in the indirect FE profiles are 27.48 kcal/mol and 28.05 kcal/mol, which deviate by only 1.18 and 1.75 kcal/mol from the direct B3LYP/MM simulation (26.30 kcal/mol).

\subsection{\label{sec:result:CPUtime}Computational expense}
The estimated wall-clock times for the computations of the direct and indirect QM/MM free energy profiles at the B3LYP/6-31G* level are listed in Table~\ref{tab:time}. For the calculations of the indirect profiles, the wall-clock time include both the time for generating the SQM trajectories and the time for the single point energy calculations at the B3LYP level. The cost at the PM3 level is similar to that at the PM6 level with the $\mathrm{S_N2}$ reaction being an exception, which was caused by the additional $d$-orbital in the basis set for the chlorine atoms in PM6. Nonetheless, both methods are orders of magnitude faster than the direct calculations of the profiles at the B3LYP level, because for each indirect profile only $10000 \times N_w$ high-level single point calculations were required, where $N_w$ is the number of windows. Because the samples for the single-point calculations at the B3LYP level were saved once every 100 MD steps in the SQM simulations in this work, the efficiency enhancement cannot exceed 100 folds. The samples can be saved every 1000 steps or even less infrequently, especially for chemical reactions with a correlation time longer than tens of picoseconds. Then, the enhancement in the efficiency will be even greater. 
\begin{table}
	\caption{\label{tab:time} Estimated wall-clock time in a unit of hours for the computations of the QM/MM free energy profiles at the B3LYP/6-31G* level. Assuming one node with 16 cores of Intel Xeon CPU E5-2660 2.20 GHz was used.}
	\newcommand{\rb}[1]{\raisebox{1.5ex}[0pt]{#1}}
	\centering
	\resizebox{\textwidth}{!}{  
		\begin{tabular}{lcccccccc}
			\hline
			&\multicolumn{8}{c}{Wall-clock Time} \\
			\cline{2-9}
			Reaction             &\multicolumn{3}{c}{indirect from PM3}&&\multicolumn{3}{c}{indirect from PM6}& \\
			\cline{2-4}\cline{6-8}
			&sampling&energy evaluation&total&&sampling&energy evaluation&total&\rb{direct$^{a}$}\\
			\hline
			BT                   & 885&1320&2205  && 890&1322&2212 & 54,900 \\
			$\mathrm{S_N2}$      & 971&1432&2403  &&2263&1480&3743 & 55,480 \\
			PT                   & 824&1282&2106  && 740&1252&1992 & 91,800 \\
			CR                   &1211&2248&3459  &&1306&2177&3483 &101,650 \\		
			\hline
		\end{tabular}
	}
	\begin{flushleft}
		\textsuperscript{\emph{a}} Using the same number of windows as in the semiempirical simulations, and one 1-ns simulation for each window.\\
		%		\textsuperscript{\emph{b}} The main chain dihedral rotation of a butane molecule.\\
		%		\textsuperscript{\emph{c}} The $\mathrm{S_{N} 2}$ reaction of \ce{CH3Cl + Cl- -> Cl- + CH3Cl}.\\
		%		\textsuperscript{\emph{d}} The glycine intramolecular proton transfer reaction.\\
		%		\textsuperscript{\emph{e}} The aliphatic Claisen rearrangement reaction of allyl vinyl ether to 4-pentenal.
	\end{flushleft}
\end{table}

\section{\label{sec:discussion}Discussion}
It is well known that the success of TP calculations heavily relies on the similarity of the reference (low-level) Hamiltonian and the target (high-level) Hamiltonian. If the important regions of these two Hamiltonians are well-separated in phase space, the important region of the target Hamiltonian cannot be sampled in finite simulation time. For those cases, the TP correction is expected to fail to yield accurate results, simply because the probability of an unsampled configuration cannot be corrected via reweighting. In this work, we only studied some simple reactions, of which the reactants have a small number of degrees-of-freedom. When the reactant becomes larger, this numerical difficulty on phase-space overlap becomes even severe, and this reweighting method is prone to more failure.
For those difficult cases, a ``case-specific'' optimization of the parameters in the reference Hamiltonian against the high-level Hamiltonian in terms of energetic and structural properties can improve the reliability of TP calculation. Kuechler et al. studied the solvation effect on the reaction mechanism of the $\mathrm{S_N2}$ reaction between chlorine anion and methyl chloride using optimized specific reaction parameters (SRP) for chlorine.\cite{KuechlerJCP2014} They showed that the optimized semi-empirical Hamiltonian can reproduce the free energy profile of a high-level method. Zhou and Pu developed a general strategy of reparametrizing semiempirical (SE) methods against \textit{ab initio} methods designated Reaction Path Force Matching (RP-FM) for QM/MM simulations in condensed phases.\cite{ZhouJCTC2014} Albeit nontransferable, RP-FM can also improve the agreement between the reference Hamiltonian and the target Hamiltonian. This topic will be covered in the next paper in this series of reports by us.

\section{\label{sec:conclusion}Conclusion}
For chemical reactions in condensed phase or enzymatic reactions, the computation of a converged free energy (FE) profile at \textit{ab initio} (\textit{ai}) QM/MM level is still far from being affordable. In this work, we proposed a new method termed MBAR+wTP to obtain the FE profile at \textit{ai} QM/MM level with much less computational expense by combining the Multistate Bennett Acceptance Ratio (MBAR) method and weighted Thermodynamic Perturbation (TP) method. This method uses the semiempirical (SE) QM/MM free energy profile obtained from MBAR analysis of the Umbrella Sampling (US) trajectories as an initial guess, followed by the SE-to-\textit{ai} correction using weighted TP. One quasi-chemical reaction and three chemical reactions are used to validate the applicability of this method. 
The SE FE profiles were found to deviate from the \textit{ai} ones by several kcal/mol in terms of the barrier height and the reaction free energy. After the SE-to-\textit{ai} correction, the FE profiles agreed much better with the direct simulated one with errors below 1 kcal/mol for most cases. To reduce the numerical difficulty in the weighted TP calculations, similar to most TP calculations, curve fitting using the Gaussian process regression  were employed to effectively reduce the fluctuations in the free energy profile. Choosing a SE method that is more similar to the \textit{ai} Hamiltonian also reduces the fluctuation and make this method more reliable. Our method is expected to fail to produce accurate reaction free energy profiles, when the SE Hamiltonian and the \textit{ai} Hamiltonian have no significant overlap in the phase space. This difficulty can be resolved by refitting the SE Hamiltonian to the \textit{ai} one via the Reaction Path Force Matching approach, which will be reported in an accompanying paper.\cite{ZhouJCTC2014} Because the samples used for the TP calculations are only one hundredth or even one thousandth of the total number of samples generated in the simulation, our method can increase the efficiency by several orders of magnitude. Since the numerical difficulty increases with the complexity of the reactions, we will validate this approach further by applying it in some real systems like enzymatic reactions.

\suppinfo
The proof that the forward and backward thermodynamic perturbation are equivalent to the Bennett Acceptance Ratio for the weighted samples from umbrella sampling, the reliability analyses of the results using overlap matrix, reweighting entropy, time interval for sampling harvesting, and the center and the force constant for each biasing window are available in the Supporting Information.

\acknowledgement
Y.M. is supported by the National Natural Science Foundation of China (Grant No. 21773066) and the Fundamental Research Funds for the Central Universities. Y.S. is grateful to Department of Energy Office of Science for support through grant DE-SC0011297. CPU time was provided by the Supercomputer Center of East China Normal University.

\clearpage

%\begin{appendices}
%\section{\label{sec:Appendix} Appendix}
%\end{appendices}
%\clearpage

%\bibliography{WHAM-TP}
\providecommand{\latin}[1]{#1}
\makeatletter
\providecommand{\doi}
{\begingroup\let\do\@makeother\dospecials
	\catcode`\{=1 \catcode`\}=2 \doi@aux}
\providecommand{\doi@aux}[1]{\endgroup\texttt{#1}}
\makeatother
\providecommand*\mcitethebibliography{\thebibliography}
\csname @ifundefined\endcsname{endmcitethebibliography}
{\let\endmcitethebibliography\endthebibliography}{}

\clearpage

\begin{tocentry}
	\centering
	\includegraphics[width=1.0\textwidth]{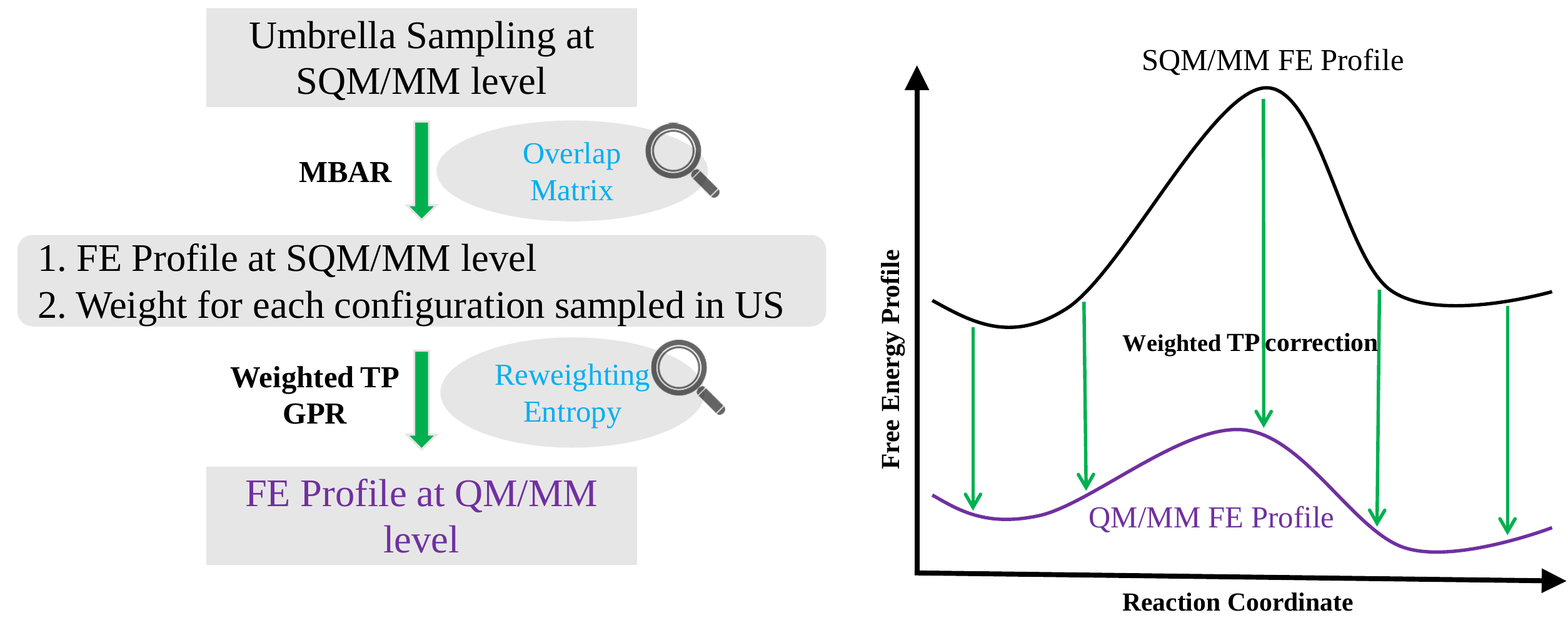}
%	\caption{Expensive \textit{ab initio} sampling for the calculations of free energy profiles can be exempted via weighted thermodynamics perturbation to the semiempirical ensemble.}
\end{tocentry}



\section*{Supplementary material}

# 1 Forward and Backward TP are equivalent to BAR for the Weighted Samples from US

Instead of calculating the weights for all the snapshots under the low-level Hamiltonian as in Eq. **??**, we can also calculate their weights under the high-level Hamiltonian as

$$w_H(\mathbf{x}_{i,l}) = \frac{e^{-\beta\left[U_H(\mathbf{x}_{i,l}) - f_H\right]}}{\sum_{k=1}^{S} N_k e^{-\beta\left[U_k^{(b)}(\mathbf{x}_{i,l}) - f_k^{(b)}\right]}}$$

$$= \frac{e^{-\beta\left[U_H(\mathbf{x}_{i,l}) - U_L(\mathbf{x}_{i,l})\right]} e^{\beta f_H}}{\sum_{k=1}^{S} N_k e^{-\beta\left[W_k(\mathbf{x}_{i,l}) - f_k^{(b)}\right]}}, \tag{S1}$$

where $e^{\beta f_H}$ is an irrelevant constant and will be canceled later in the normalization. With these weights, the free energy difference can be obtained via a TP calculation from the high-level Hamiltonian to the low-level Hamiltonian by

$$\Delta F(\eta) = \frac{1}{\beta} \ln \frac{\sum_{i=1}^{S}\sum_{l=1}^{N_i} w_H(\mathbf{x}_\eta) \delta\left(\eta_{(i,l)} - \eta\right) e^{-\beta[U_L(\mathbf{x}_\eta) - U_H(\mathbf{x}_\eta)]}}{\sum_{i=1}^{S}\sum_{l=1}^{N_i} w_H(\mathbf{x}_\eta) \delta\left(\eta_{(i,l)} - \eta\right)}$$

$$= \frac{1}{\beta} \ln \frac{\sum_{i=1}^{S}\sum_{l=1}^{N_i} w_L(\mathbf{x}_\eta) \delta\left(\eta_{(i,l)} - \eta\right)}{\sum_{i=1}^{S}\sum_{l=1}^{N_i} w_L(\mathbf{x}_\eta) \delta\left(\eta_{(i,l)} - \eta\right) e^{-\beta\left[U_H\left(\mathbf{x}_{\eta_{(i,l)}}\right) - U_L\left(\mathbf{x}_{\eta_{(i,l)}}\right)\right]}}$$

$$= -\frac{1}{\beta} \ln \frac{\sum_{i=1}^{S}\sum_{l=1}^{N_i} w_L\left(\mathbf{x}_{\eta_{(i,l)}}\right) e^{-\beta\left[U_H\left(\mathbf{x}_{\eta_{(i,l)}}\right) - U_L\left(\mathbf{x}_{\eta_{(i,l)}}\right)\right]}}{\sum_{i=1}^{S}\sum_{l=1}^{N_i} w_L\left(\mathbf{x}_{\eta_{(i,l)}}\right)}, \tag{S2}$$

which is exactly Eq. **??**. We can also use Bennett acceptance ratio instead of TP via

$$\Delta F(\eta) = -\beta^{-1} \ln \left[\frac{\langle f\left(U_H(\mathbf{x}_\eta) - U_L(\mathbf{x}_\eta) - C\right)\rangle_L}{\langle f\left(U_L(\mathbf{x}_\eta) - U_H(\mathbf{x}_\eta) + C\right)\rangle_H} \exp\left(-\beta C\right)\right], \tag{S3}$$

where $f(x) = \frac{1}{1+\exp(\beta x)}$ is the Fermi function and $C = \beta^{-1} \ln \frac{Q_L}{Q_H}$ with $Q_L$ and $Q_H$ being the partition function under the low-level and the high-level Hamiltonians, respectively. We can expand the ensemble averages of the Fermi functions as

$$\langle f(U_H(\mathbf{x}_\eta) - U_L(\mathbf{x}_\eta) - C)\rangle_L = \frac{\sum_{i=1}^{S}\sum_{l=1}^{N_i} w_L(\mathbf{x}_\eta) \delta(\eta_{(i,l)} - \eta) \frac{1}{1+\exp[\beta(U_H(\mathbf{x}_\eta) - U_L(\mathbf{x}_\eta) - C)]}}{\sum_{i=1}^{S}\sum_{l=1}^{N_i} w_L(\mathbf{x}_\eta) \delta(\eta_{(i,l)} - \eta)} \quad (S4)$$

and

$$\begin{aligned}
\langle f(U_L(\mathbf{x}_\eta) - U_H(\mathbf{x}_\eta) + C)\rangle_H &= \frac{\sum_{i=1}^{S}\sum_{l=1}^{N_i} w_H(\mathbf{x}_\eta) \delta(\eta_{(i,l)} - \eta) \frac{1}{1+\exp[\beta(U_L(\mathbf{x}_\eta) - U_H(\mathbf{x}_\eta) + C)]}}{\sum_{i=1}^{S}\sum_{l=1}^{N_i} w_H(\mathbf{x}_\eta) \delta(\eta_{(i,l)} - \eta)} \\
&= \frac{\sum_{i=1}^{S}\sum_{l=1}^{N_i} w_H(\mathbf{x}_\eta) \delta(\eta_{(i,l)} - \eta) \frac{\exp[\beta(U_H(\mathbf{x}_\eta) - U_L(\mathbf{x}_\eta) - C)]}{1+\exp[\beta(U_H(\mathbf{x}_\eta) - U_L(\mathbf{x}_\eta) - C)]}}{\sum_{i=1}^{S}\sum_{l=1}^{N_i} w_H(\mathbf{x}_\eta) \delta(\eta_{(i,l)} - \eta)} \\
&= \frac{\sum_{i=1}^{S}\sum_{l=1}^{N_i} w_L(\mathbf{x}_\eta) \delta(\eta_{(i,l)} - \eta) \frac{\exp(-\beta C)}{1+\exp[\beta(U_H(\mathbf{x}_\eta) - U_L(\mathbf{x}_\eta) - C)]}}{\sum_{i=1}^{S}\sum_{l=1}^{N_i} w_L(\mathbf{x}_\eta) e^{-\beta[U_H(\mathbf{x}_{i,l}) - U_L(\mathbf{x}_{i,l})]} \delta(\eta_{(i,l)} - \eta)}
\end{aligned} \quad (S5)$$

Taking Eq. S4 and S5 into Eq. S3, we find

$$\begin{aligned}
\Delta F(\eta) &= -\beta^{-1} \ln \left[ \frac{\frac{\sum_{i=1}^{S}\sum_{l=1}^{N_i} w_L(\mathbf{x}_\eta)\delta(\eta_{(i,l)}-\eta)\frac{1}{1+\exp[\beta(U_H(\mathbf{x}_\eta)-U_L(\mathbf{x}_\eta)-C)]}}{\sum_{i=1}^{S}\sum_{l=1}^{N_i} w_L(\mathbf{x}_\eta)\delta(\eta_{(i,l)}-\eta)}}{\frac{\sum_{i=1}^{S}\sum_{l=1}^{N_i} w_L(\mathbf{x}_\eta)\delta(\eta_{(i,l)}-\eta)\frac{\exp(-\beta C)}{1+\exp[\beta(U_H(\mathbf{x}_\eta)-U_L(\mathbf{x}_\eta)-C)]}}{\sum_{i=1}^{S}\sum_{l=1}^{N_i} w_L(\mathbf{x}_\eta)e^{-\beta[U_H(\mathbf{x}_{i,l})-U_L(\mathbf{x}_{i,l})]}\delta(\eta_{(i,l)}-\eta)}} \exp(-\beta C) \right] \\
&= -\frac{1}{\beta} \ln \frac{\sum_{i=1}^{S}\sum_{l=1}^{N_i} w_L\left(\mathbf{x}_{\eta_{(i,l)}}\right) e^{-\beta\left[U_H\left(\mathbf{x}_{\eta_{(i,l)}}\right) - U_L\left(\mathbf{x}_{\eta_{(i,l)}}\right)\right]}}{\sum_{i=1}^{S}\sum_{l=1}^{N_i} w_L\left(\mathbf{x}_{\eta_{(i,l)}}\right)},
\end{aligned} \quad (S6)$$

which is consistent with the TP calculations. These results are not surprising, because these two calculations have the same amount of information as the forward TP shown in the main text.

## 2  Reliability of the MBAR Analysis of the Trajectories from Umbrella Samplings

The reliabilities of the MBAR calculations are examined by calculating the overlap matrix proposed by Mobley et. al.,[1] which essentially measures the magnitude of the phase space overlap. With the weight of the $l$th configuration in the $i$th biased simulation appearing in the $t$th simulation defined as

$$w_t(\mathbf{x}_{i,l}) = \frac{e^{-\beta\left[W_t(\mathbf{x}_{i,l}) - f_t^{(b)}\right]}}{\sum_{k=1}^{S} N_k e^{-\beta\left[W_k(\mathbf{x}_{i,l}) - f_k^{(b)}\right]}}, \tag{S7}$$

the elements of the $S \times S$ overlap matrix are[1]

$$\begin{aligned}
O_{tt'} &= \sum_{i=1}^{S} \sum_{l=1}^{N_i} N_t w_t(\mathbf{x}_{i,l}) w_{t'}(\mathbf{x}_{i,l}) \\
&= \sum_{i=1}^{S} \sum_{l=1}^{N_i} \frac{N_t e^{-\beta\left[W_t(\mathbf{x}_{i,l}) - f_t^{(b)}\right]} e^{-\beta\left[W_{t'}(\mathbf{x}_{i,l}) - f_{t'}^{(b)}\right]}}{\left\{\sum_{k=1}^{S} N_k e^{-\beta\left[W_k(\mathbf{x}_{i,l}) - f_k^{(b)}\right]}\right\}^2}.
\end{aligned} \tag{S8}$$

Consecutive windows should have substantial overlap with the diagonal and the first off-diagonal elements no smaller than 0.03 as recommended.[1]

The US simulations were performed at the PM3 and PM6 and B3LYP levels, from which the FE profiles were calculated by the MBAR method. In order to assess whether the phase space overlap between each pair of the neighboring windows was sufficient, the overlap matrices were calculated for these trajectories. As shown in Fig. S1 for the dihedral rotation of butane, all elements in the diagonal and the first off-diagonals of all matrices are larger than 0.22. As shown in Fig. S4 for the $S_N2$ reaction, all elements in the diagonal and the first off-diagonals of all matrices are larger than 0.1. As shown in Fig. S7 for the proton transfer in glycine, all elements in the diagonal and the first off-diagonals of all matrices are all larger than 0.07. As shown in Fig. S10, all elements in the diagonal and the first off-diagonals of all

matrices are no smaller than 0.09. All of them are larger than the lower-limit 0.03, indicating that the phase space overlap is sufficient for the subsequent MBAR analysis. Therefore, the SQM and QM/MM FE profiles calculated by the MBAR method are statistically reliable.

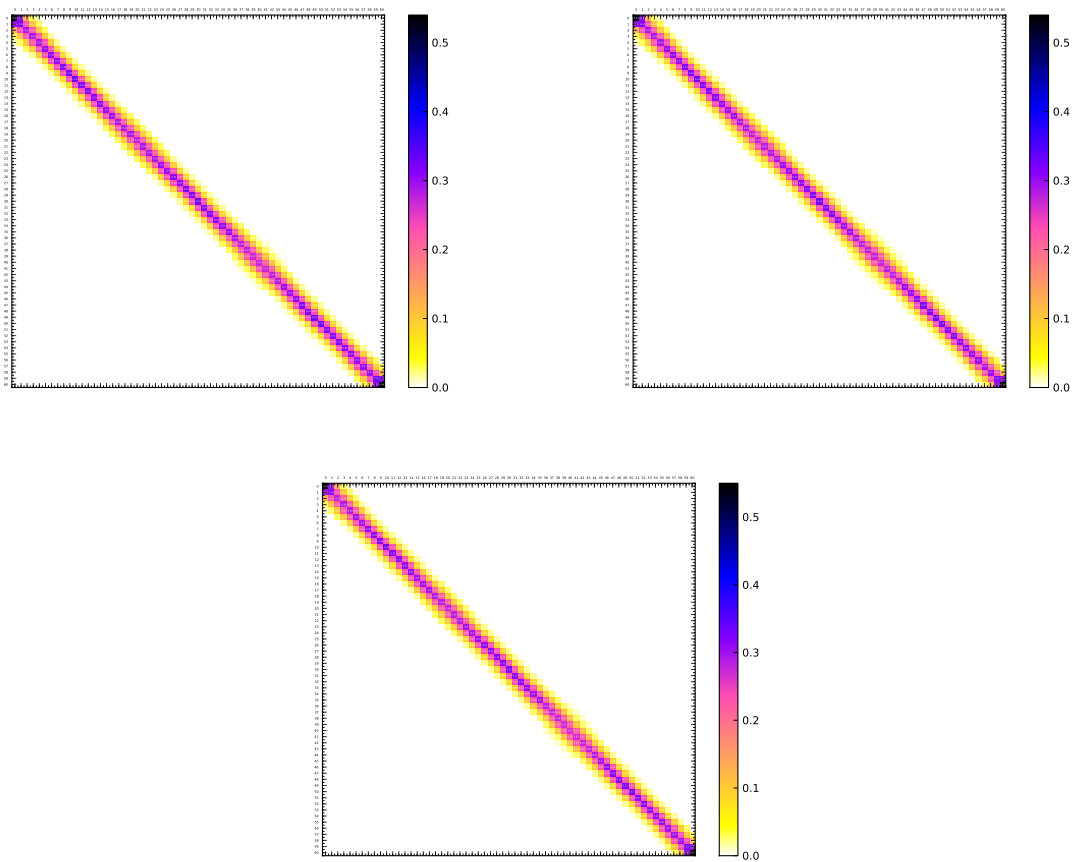

Figure S1: The overlap matrices defined by Eq. S8 for the US simulations at the PM3 level (upper left) and the PM6 level (upper right) and the B3LYP level (bottom) for the main chain dihedral rotation of a butane molecule. Both axes are the simulation windows.

6

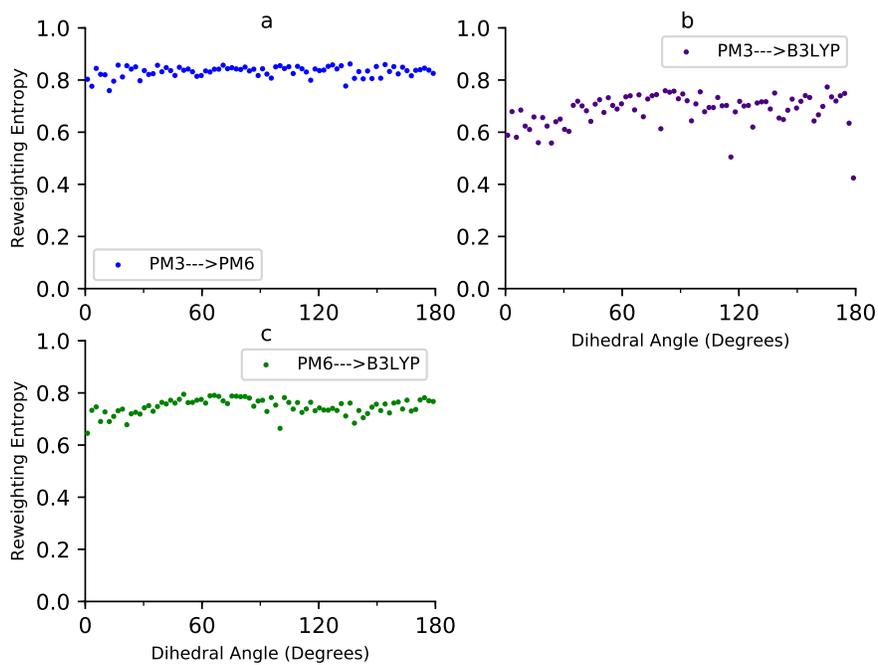

Figure S2: The reweighting entropies in the weighted TP calculations for the main chain dihedral rotation of a butane molecule.

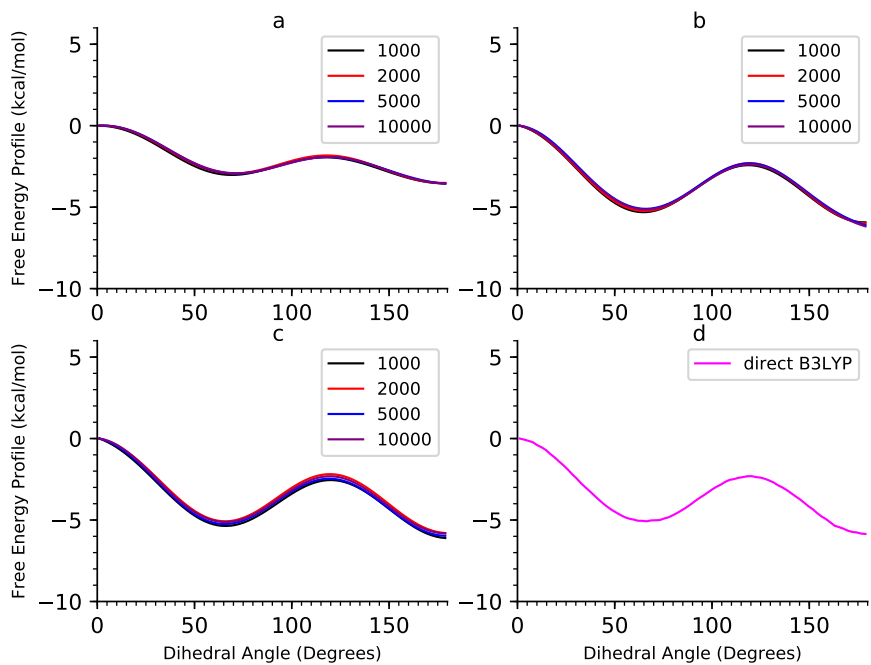

Figure S3: The indirect FE profiles for the main chain dihedral rotation of a butane molecule (a) at PM6 level by reweighting from PM3, (b) at B3LYP level by reweighting from PM3, (c) at B3LYP level by reweighting from PM6 as well as (d) the direct FE profile at the B3LYP level (magenta). In subfigures a, b and c, there are four kinds of different harvested time 0.1 ps (10000 frames of configurations, purple), 0.2 ps (5000 frames of configurations, blue), 0.5 ps (2000 frames of configurations, red) and 1 ps (1000 frames of configurations, black).

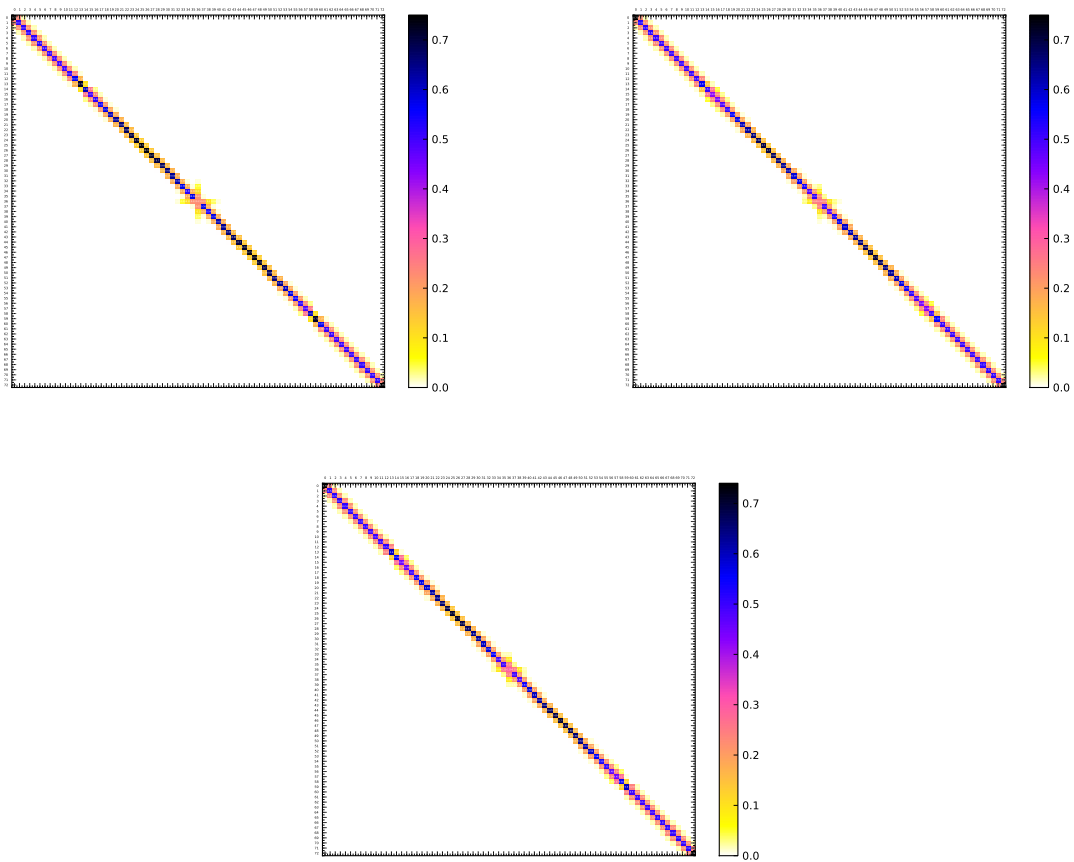

Figure S4: The overlap matrices defined by Eq. S8 for the US simulations at the PM3 level (upper left) and the PM6 level (upper right) and the B3LYP level (bottom) for the $S_N 2$ reaction of $CH_3Cl + Cl^- \longrightarrow Cl^- + CH_3Cl$. Both axes are the simulation windows.

9

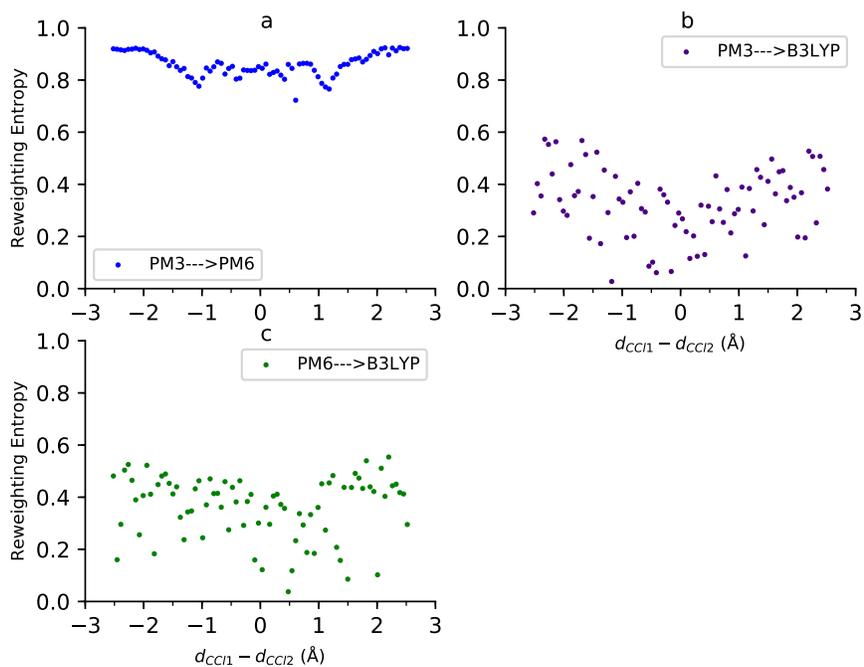

Figure S5: The reweighting entropies in the weighted TP calculations for the $S_N 2$ reaction of $CH_3Cl + Cl^- \longrightarrow Cl^- + CH_3Cl$.

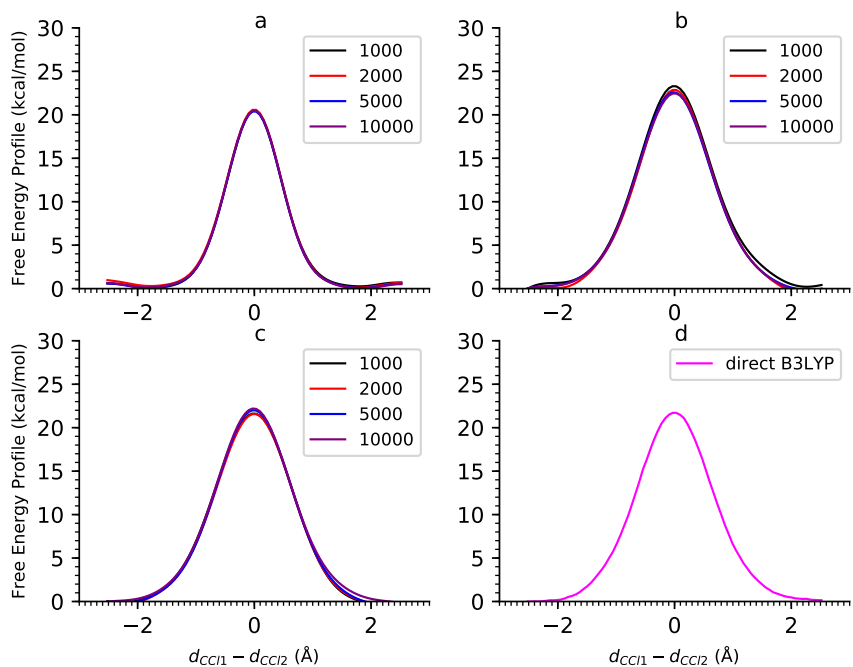

Figure S6: The indirect FE profiles for the SN2 reaction of $CH_3Cl + Cl^- \longrightarrow Cl^- + CH_3Cl$ (a) at PM6 level by reweighting from PM3, (b) at B3LYP level by reweighting from PM3, (c) at B3LYP level by reweighting from PM6 as well as (d) the direct FE profile at the B3LYP level (magenta). In subfigures a, b and c, there are four kinds of different harvested time 0.1 ps (10000 frames of configurations, purple), 0.2 ps (5000 frames of configurations, blue), 0.5 ps (2000 frames of configurations, red) and 1 ps (1000 frames of configurations, black).

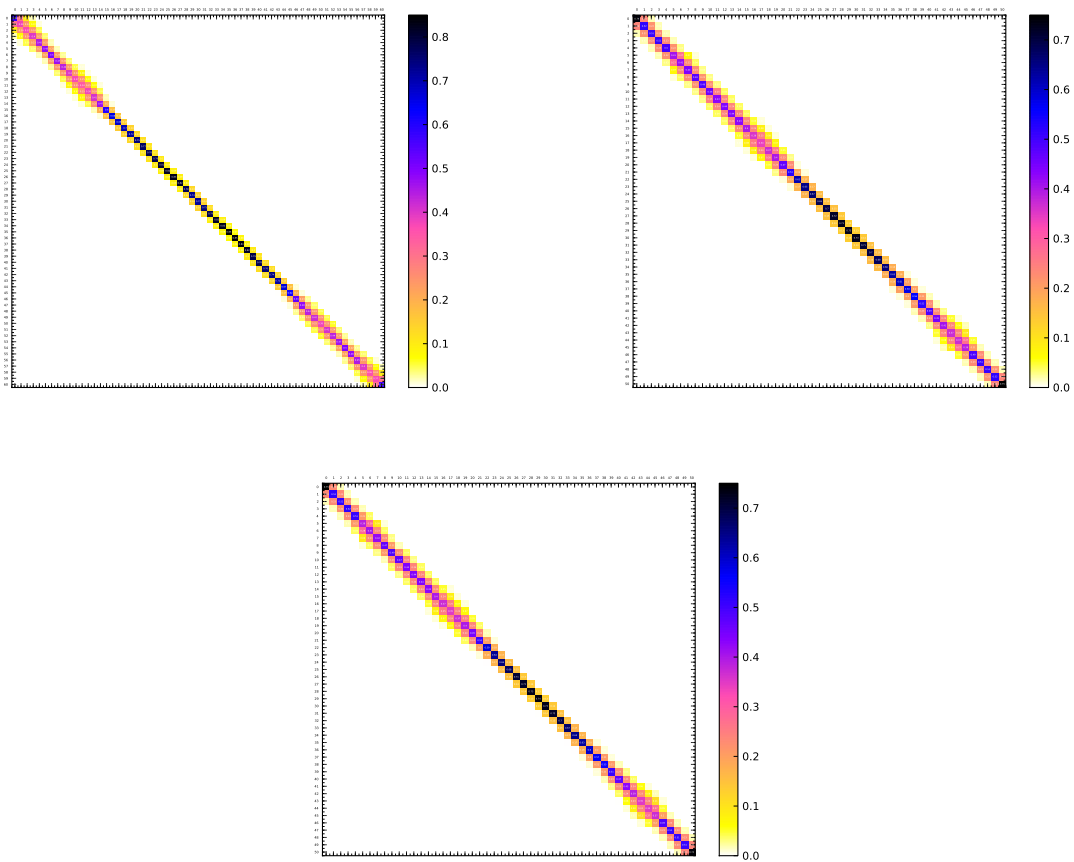

Figure S7: The overlap matrices defined by Eq. S8 for the US simulations at the PM3 level (upper left) and the PM6 level (upper right) and the B3LYP level (bottom) for the glycine intramolecular proton transfer reaction. Both axes are the simulation windows.

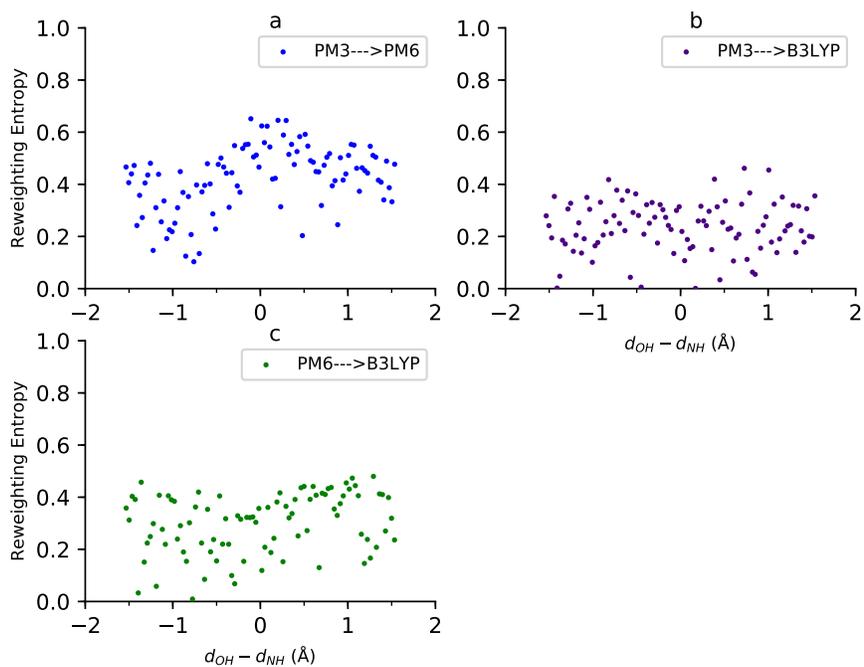

Figure S8: The reweighting entropies in the weighted TP calculations for the glycine intramolecular proton transfer reaction.

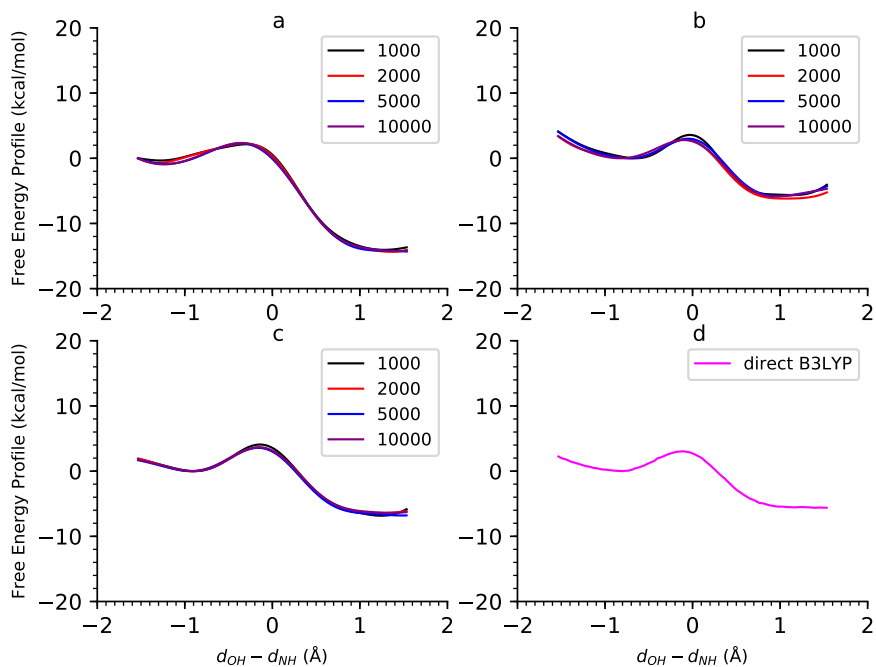

Figure S9: The indirect FE profiles for the glycine intramolecular proton transfer reaction (a) at PM6 level by reweighting from PM3, (b) at B3LYP level by reweighting from PM3, (c) at B3LYP level by reweighting from PM6 as well as (d) the direct FE profile at the B3LYP level (magenta). In subfigures a, b and c, there are four kinds of different harvested time 0.1 ps (10000 frames of configurations, purple), 0.2 ps (5000 frames of configurations, blue), 0.5 ps (2000 frames of configurations, red) and 1 ps (1000 frames of configurations, black).

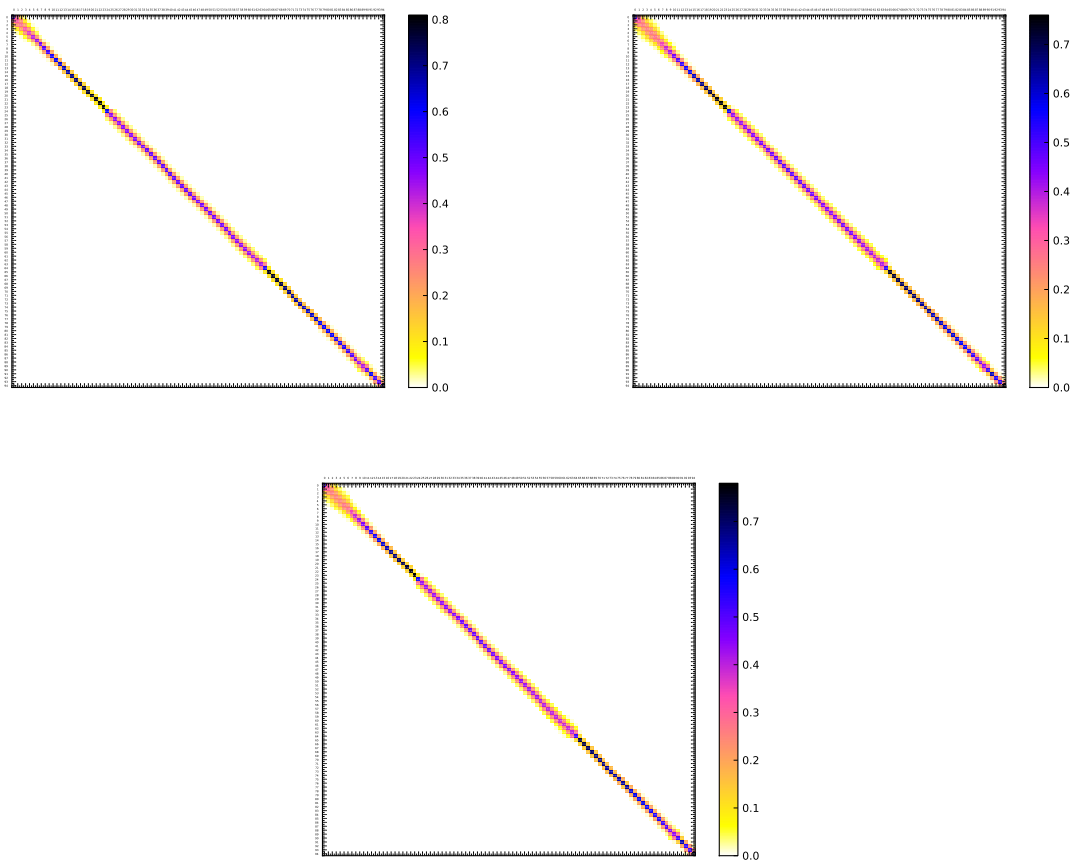

Figure S10: The overlap matrices defined by Eq. S8 for the US simulations at the PM3 level (upper left) and the PM6 level (upper right) and the B3LYP level (bottom) for the aliphatic Claisen rearrangement reaction of allyl vinyl ether (AVE) to 4-pentenal. Both axes are the simulation windows.

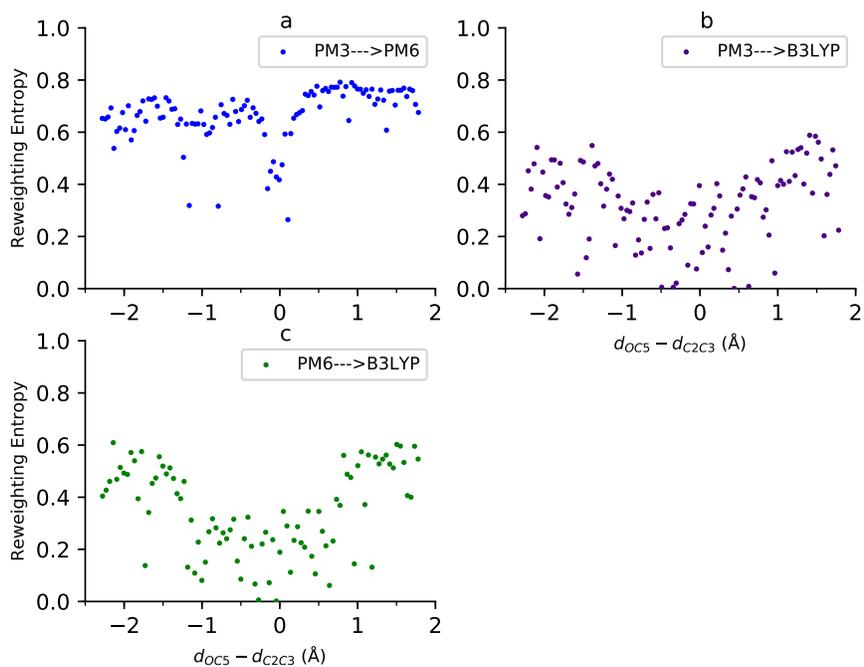

Figure S11: The reweighting entropies in the weighted TP calculations for the aliphatic Claisen rearrangement reaction of allyl vinyl ether (AVE) to 4-pentenal.

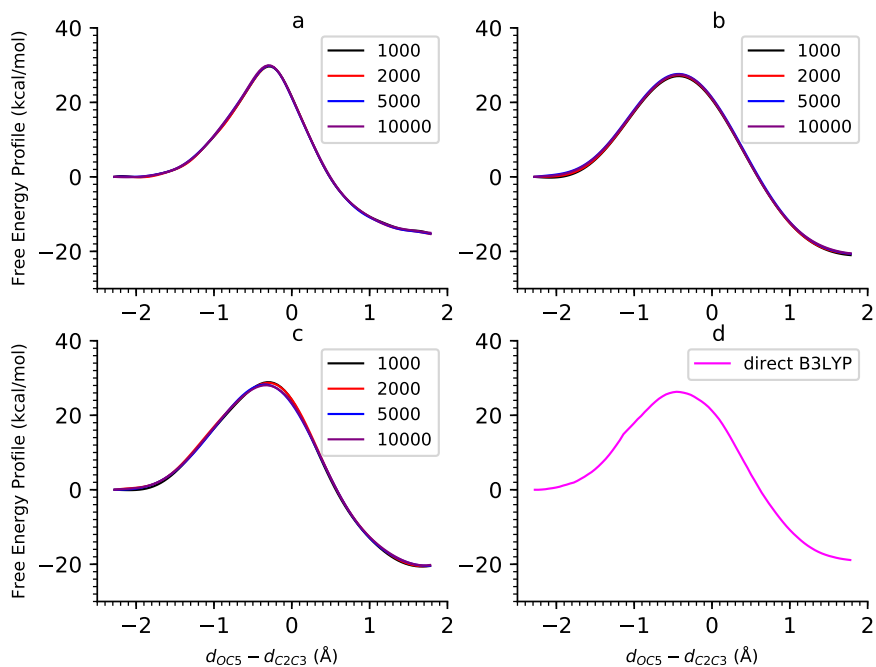

Figure S12: The indirect FE profiles for the aliphatic Claisen rearrangement reaction of allyl vinyl ether (AVE) to 4-pentenal (a) at PM6 level by reweighting from PM3, (b) at B3LYP level by reweighting from PM3, (c) at B3LYP level by reweighting from PM6 as well as (d) the direct FE profile at the B3LYP level (magenta). In subfigures a, b and c, there are four kinds of different harvested time 0.1 ps (10000 frames of configurations, purple), 0.2 ps (5000 frames of configurations, blue), 0.5 ps (2000 frames of configurations, red) and 1 ps (1000 frames of configurations, black).

# 3 The center $\eta_i$ and the force constant $k_i$ of the biasing potential in the US simulations

## 3.1 The main chain dihedral rotation of a butane molecule

There are 61 US windows at PM3, PM6 and B3LYP levels, of which (centers of restraint (in degree), force constant (in $kcal/mol/\deg^2$)) are:

(0, 100), (3, 100), (6, 110), (9, 115), (12, 120), (15, 125), (18, 130), (21, 135), (24, 140), (27, 145), (30, 150), (33, 145), (36, 140), (39, 135), (42, 130), (45, 125), (48, 120), (51, 115), (54, 110), (57, 105), (60, 100), (63, 105), (66, 110), (69, 115), (72, 120), (75, 125), (78, 130), (81, 135), (84, 140), (87, 145), (90, 150), (93, 145), (96, 140), (99, 135), (102, 130), (105, 125), (108, 120), (111, 115), (114, 110), (117, 105), (120, 100), (123, 105), (126, 110), (129, 115), (132, 120), (135, 125), (138, 130), (141, 135), (144, 140), (147, 145), (150, 150), (153, 145), (156, 140), (159, 135), (162, 130), (165, 125), (168, 120), (171, 115), (174, 110), (177, 105), (180, 100).

## 3.2 The $S_N 2$ reaction

There are 73 US windows at PM3 level, of which (center of restraint (in Å), force constant (in $kcal/mol/\text{Å}^2$)) are:

(-2.5, 100), (-2.4, 100), (-2.3, 100), (-2.2, 100), (-2.1, 100), (-2.0, 100), (-1.9, 100), (-1.8, 100), (-1.7, 100), (-1.6, 100), (-1.5, 100), (-1.4, 100), (-1.3, 100), (-1.2, 200), (-1.1, 300), (-1.05, 350), (-1.0, 400), (-0.95, 450), (-0.9, 500), (-0.85, 550), (-0.8, 600), (-0.75, 650), (-0.7, 700), (-0.65, 750), (-0.6, 800), (-0.55, 850), (-0.5, 900), (-0.45, 850), (-0.4, 800), (-0.35, 750), (-0.3, 700), (-0.25, 650), (-0.2, 600), (-0.15, 550), (-0.1, 500), (-0.05, 450), (0.0, 100), (0.05, 450), (0.1, 500), (0.15, 550), (0.2, 600), (0.25, 650), (0.3, 700), (0.35, 750), (0.4, 800), (0.45, 850), (0.5, 900), (0.55, 850), (0.6, 800), (0.65, 750), (0.7, 700), (0.75, 650), (0.8, 600), (0.85, 550), (0.9, 500), (0.95, 450), (1.0, 400), (1.05, 350), (1.1, 300), (1.2, 200), (1.3, 100), (1.4,

100), (1.5, 100), (1.6, 100), (1.7, 100), (1.8, 100), (1.9, 100), (2.0, 100), (2.1, 100), (2.2, 100), (2.3, 100), (2.4, 100), (2.5, 100).

There are 73 US windows at PM6 and B3LYP levels, of which (center of restraint (in Å), force constant (in $kcal/mol/\text{Å}^2$)) are:

(-2.5, 100), (-2.4, 100), (-2.3, 100), (-2.2, 100), (-2.1, 100), (-2.0, 100), (-1.9, 100), (-1.8, 100), (-1.7, 100), (-1.6, 100), (-1.5, 100), (-1.4, 100), (-1.3, 100), (-1.2, 100), (-1.1, 200), (-1.05, 250), (-1.0, 300), (-0.95, 350), (-0.9, 400), (-0.85, 450), (-0.8, 500), (-0.75, 550), (-0.7, 600), (-0.65, 650), (-0.6, 700), (-0.55, 750), (-0.5, 800), (-0.45, 750), (-0.4, 700), (-0.35, 650), (-0.3, 600), (-0.25, 550), (-0.2, 500), (-0.15, 450), (-0.1, 400), (-0.05, 350), (0.0, 100), (0.05, 350), (0.1, 400), (0.15, 450), (0.2, 500), (0.25, 550), (0.3, 600), (0.35, 650), (0.4, 700), (0.45, 750), (0.5, 800), (0.55, 750), (0.6, 700), (0.65, 650), (0.7, 600), (0.75, 550), (0.8, 500), (0.85, 450), (0.9, 400), (0.95, 350), (1.0, 300), (1.05, 250), (1.1, 200), (1.2, 100), (1.3, 100), (1.4, 100), (1.5, 100), (1.6, 100), (1.7, 100), (1.8, 100), (1.9, 100), (2.0, 100), (2.1, 100), (2.2, 100), (2.3, 100), (2.4, 100), (2.5, 100).

### 3.3 The glycine intramolecular proton transfer reaction

There are 61 US windows at PM3 level, of which (center of restraint (in Å), force constant (in $kcal/mol/\text{Å}^2$)) are:

(-1.5, 100), (-1.45, 150), (-1.4, 200), (-1.35, 250), (-1.3, 300), (-1.25, 350), (-1.2, 400), (-1.15, 350), (-1.1, 300), (-1.05, 250), (-1.0, 200), (-0.95, 250), (-0.9, 300), (-0.85, 350), (-0.8, 400), (-0.75,550), (-0.7, 700), (-0.65, 800), (-0.6, 900), (-0.55, 950), (-0.5, 1000), (-0.45, 1050), (-0.4, 1100), (-0.35, 1250), (-0.3, 1300), (-0.25, 1350), (-0.2, 1300), (-0.15, 1150), (-0.1, 1050), (-0.05, 800), (0.0,700), (0.05, 800), (0.1, 1100), (0.15, 1150), (0.2, 1200), (0.25, 1150), (0.3, 1100), (0.35, 1050), (0.4, 1000), (0.45, 950), (0.5, 900), (0.55, 850), (0.6, 800), (0.65, 750), (0.7, 700), (0.75, 550), (0.8, 300), (0.85, 250), (0.9, 200), (0.95, 150), (1.0, 200), (1.05, 250), (1.1, 300), (1.15, 350), (1.2, 400), (1.25, 350), (1.3, 300), (1.35, 250), (1.4, 200), (1.45, 150), (1.5, 100).

There are 51 US windows at PM6 and B3LYP levels, of which (center of restraint (in Å), force constant (in $kcal/mol/\text{Å}^2$)) are:

(-1.5, 100), (-1.4, 100), (-1.3, 100), (-1.2, 100), (-1.1, 100), (-1.0, 100), (-0.95, 150), (-0.9, 200), (-0.85, 250), (-0.8, 300), (-0.75, 350), (-0.7, 400), (-0.65, 450), (-0.6, 500), (-0.55, 550), (-0.5, 600), (-0.45, 650), (-0.4, 700), (-0.35, 750), (-0.3, 800), (-0.25, 850), (-0.2, 900), (-0.15, 850), (-0.1, 800), (-0.05, 750), (0.0, 700), (0.05, 650), (0.1, 600), (0.15, 450), (0.2, 300), (0.25, 250), (0.3, 200), (0.35, 200), (0.4, 200), (0.45, 250), (0.5, 300), (0.55, 300), (0.6, 300), (0.65, 300), (0.7, 300), (0.75, 350), (0.8, 400), (0.85, 350), (0.9, 300), (0.95, 200), (1.0, 100), (1.1, 100), (1.2, 100), (1.3, 100), (1.4, 100), (1.5, 100).

## 3.4 The Claisen rearrangement reaction

There are 95 US windows at PM3 level, of which (center of restraint (in Å), force constant (in $kcal/mol/\text{Å}^2$)) are:

(-2.2, 100), (-2.1, 100), (-2.0, 100), (-1.9, 150), (-1.8, 200), (-1.75, 250), (-1.7, 300), (-1.65, 350), (-1.6, 400), (-1.55, 450), (-1.5, 500), (-1.45, 500), (-1.4, 500), (-1.35, 550), (-1.3, 550), (-1.25, 550), (-1.2, 550), (-1.15, 600), (-1.1, 650), (-1.05, 650), (-1.0, 650), (-0.95, 700), (-0.9, 700), (-0.85, 750), (-0.8, 750), (-0.75, 800), (-0.7, 850), (-0.65, 950), (-0.6, 1000), (-0.55, 1000), (-0.5, 1050), (-0.475, 1050), (-0.45, 1050), (-0.425, 1075), (-0.4, 1100), (-0.375, 1125), (-0.35, 1150), (-0.325, 1175), (-0.3, 1200), (-0.275, 1225), (-0.25, 1250), (-0.225, 1275), (-0.2, 1300), (-0.175, 1325), (-0.15, 1350), (-0.125, 1375), (-0.1, 1400), (-0.075, 1425), (-0.05, 1450), (-0.025, 1475), (0.0, 1500), (0.025, 1525), (0.05, 1550), (0.075, 1575), (0.1, 1600), (0.125, 1575), (0.15, 1550), (0.175, 1525), (0.2, 1500), (0.225, 1475), (0.25, 1450), (0.275, 1425), (0.3, 1400), (0.325, 1375), (0.35, 1350), (0.375, 1325), (0.4, 1300), (0.425, 1275), (0.45, 1250), (0.475, 1225), (0.5, 1200), (0.55, 1150), (0.6, 1100), (0.65, 1050), (0.7, 1000), (0.75, 950), (0.8, 900), (0.85, 850), (0.9, 800), (0.95, 750), (1.0, 700), (1.05, 650), (1.1, 600), (1.15, 550), (1.2, 500), (1.25, 450), (1.3, 400), (1.35, 350), (1.4, 300), (1.45, 250), (1.5, 200), (1.55, 150), (1.6, 100), (1.65, 100), (1.7, 100).

There are 95 US windows at PM6 and B3LYP levels, of which (center of restraint (in Å), force constant (in $kcal/mol/Å^2$)) are:

(-2.2, 100), (-2.1, 100), (-2.0, 100), (-1.9, 150), (-1.8, 200), (-1.75, 250), (-1.7, 300), (-1.65, 350), (-1.6, 400), (-1.55, 450), (-1.5, 500), (-1.45, 500), (-1.4, 500), (-1.35, 550), (-1.3, 550), (-1.25, 550), (-1.2, 550), (-1.15, 600), (-1.1, 600), (-1.05, 600), (-1.0, 650), (-0.95, 650), (-0.9, 650), (-0.85, 700), (-0.8, 700), (-0.75, 700), (-0.7, 750), (-0.65, 750), (-0.6, 800), (-0.55, 800), (-0.5, 850), (-0.475, 850), (-0.45, 850), (-0.425, 875), (-0.4, 900), (-0.375, 925), (-0.35, 950), (-0.325, 975), (-0.3, 1000), (-0.275, 1025), (-0.25, 1050), (-0.225, 1075), (-0.2, 1100), (-0.175, 1125), (-0.15, 1150), (-0.125, 1175), (-0.1, 1200), (-0.075, 1225), (-0.05, 1250), (-0.025, 1275), (0.0, 1300), (0.025, 1325), (0.05, 1350), (0.075, 1375), (0.1, 1400), (0.125, 1375), (0.15, 1350), (0.175, 1325), (0.2, 1300), (0.225, 1275), (0.25, 1250), (0.275, 1225), (0.3, 1200), (0.325, 1175), (0.35, 1150), (0.375, 1125), (0.4, 1100), (0.425, 1075), (0.45, 1050), (0.475, 1025), (0.5, 1000), (0.55, 950), (0.6, 900), (0.65, 850), (0.7, 800), (0.75, 750), (0.8, 700), (0.85, 650), (0.9, 600), (0.95, 550), (1.0, 500), (1.05, 450), (1.1, 400), (1.15, 350), (1.2, 300), (1.25, 250), (1.3, 200), (1.35, 150), (1.4, 100), (1.45, 100), (1.5, 100), (1.55, 100), (1.6, 100), (1.65, 100), (1.7, 100).

# Graphical TOC Entry

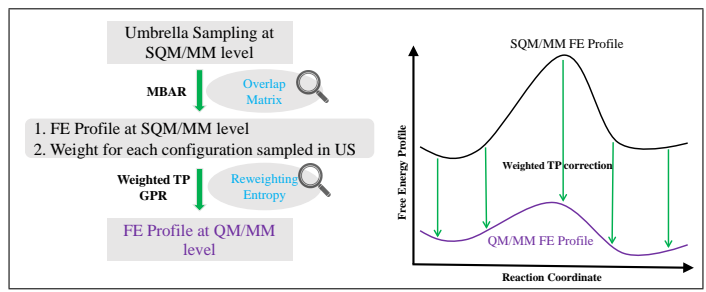


\end{document}